\documentclass[aps,prd,preprint,nofootinbib,titlepage,amsmath,amsfonts]{revtex4-2}
\usepackage[colorlinks,allcolors=blue]{hyperref}
\usepackage[T1]{fontenc}
\usepackage[USenglish]{babel}
\usepackage{bm}
\usepackage{bbm}
\usepackage{enumerate}
\usepackage{multirow}
\usepackage{microtype}
\usepackage{graphicx}
\newcommand{\cg}[6]{
\biggl\langle
\begin{matrix}
#1	& #3	\\
#2	& #4
\end{matrix}
\bigg\vert
\begin{matrix}
#5	\\
#6 
\end{matrix}
\biggr\rangle
}

\newcommand{\smallcg}[6]{
\bigl\langle
\begin{smallmatrix}
#1	& #3	\\
#2	& #4	\\
\end{smallmatrix}
\big\vert
\begin{smallmatrix}
#5	\\
#6 	\\
\end{smallmatrix}
\bigr\rangle
}

\newcommand{\vket}[4]{
\biggl\lvert
#1
\begin{matrix}
#2	\\
#3
\end{matrix}
#4
\biggr\rangle
}

\newcommand{\smallvket}[4]{
\Bigl\lvert
#1
\begin{smallmatrix}
#2	\\
#3
\end{smallmatrix}
#4
\Bigr\rangle
}

\begin{document}

\title{Model-independent predictions for decays\texorpdfstring{\\}{ }of double-heavy hadrons into pairs of heavy hadrons}
\author{R. Bruschini}
\email{bruschini.1@osu.edu}
\affiliation{Department of Physics, The Ohio State University, Columbus, OH 43210, USA}

\begin{abstract}
Double-heavy hadrons can decay into pairs of heavy hadrons through transitions from confining Born-Oppenheimer potentials to heavy-hadron-pair potentials with the same Born-Oppenheimer quantum numbers. The states of the double-heavy hadron are constrained by a Born-Oppenheimer exclusion principle from the identical heavy quarks. The states of a pair of identical heavy hadrons are constrained by exclusion principles from identical particles. The transitions are also constrained by conservation of angular momentum and parity. From these constraints, we derive model-independent selection rules for decays of double-heavy hadrons into pairs of heavy hadrons. The coupling potentials are expressed as sums of products of Born-Oppenheimer transition amplitudes and angular-momentum coefficients. If there is a single dominant Born-Oppenheimer transition amplitude, it factors out of the coupling potentials between double-heavy hadrons in the same Born-Oppenheimer multiplet and pairs of heavy hadrons in specific heavy-quark-spin-symmetry multiplets and out of the corresponding partial decay rates. As examples, we discuss the Born-Oppenheimer potentials and multiplets for conventional double-heavy baryons and for double-heavy tetraquark mesons. We also discuss the relative partial decay rates for conventional double-heavy baryons into pairs of heavy hadrons.
\end{abstract}

\maketitle

\section{Introduction}

For more than 20 years, the discoveries of dozens of exotic heavy hadrons have laid bare a serious gap in our understanding of strong interactions \cite{Leb17,Bra20,Che23}. Many theoretical studies have been dedicated to exotic \emph{hidden-heavy} hadrons, which feature several well-established $\mathit{XYZ}$ hadrons. On the other hand, the only well-established \emph{double-heavy} hadrons are the conventional double-charm baryon $\Xi_c^{++}$ \cite{Aai17} and the exotic double-charm tetraquark meson $T_{cc}^+$ \cite{Aai22}, both of which were discovered by the LHCb Collaboration. Despite the limited data, exotic double-heavy hadrons have been subject to intense theoretical scrutiny. In particular, double-heavy tetraquark mesons have been studied using lattice QCD \cite{Bro12,Bic16,Bic17,Fra17,Jun19,Les19,Pad22}, effective field theories \cite{Flem20,Fei21,Braa22,Du22}, heavy-quark symmetry \cite{Eic17,Kar17}, and models; see, for instance, the recent Refs.~\cite{Men21,San22,Mai23,Leb24,Ma24,Ort24} and references therein.

Here we focus on double-heavy hadrons following another approach that is firmly based on the fundamental theory of strong interactions: the \emph{Born-Oppenheimer (B\nobreakdash-O) approximation} for QCD \cite{Jug99}. In this approximation, the calculation of the double-heavy-hadron spectrum amounts to a two-step procedure. The first step is to calculate \emph{B\nobreakdash-O potentials} using lattice QCD with two static color sources. The second step is to take into account the heavy-quark motion by solving a coupled-channel Schr\"odinger equation in the B\nobreakdash-O potentials. It has been shown that this approach is equivalent to the leading order in a rigorous effective theory of QCD \cite{Ber15,On17,Bra18a,Sot20a}. The B\nobreakdash-O approximation has been applied to double-heavy baryons \cite{Sot20b} and to double-heavy tetraquark mesons \cite{Bro12,Bic16,Bic17}. The philosophy of the B\nobreakdash-O approximation has also inspired constituent models of double-heavy tetraquark mesons \cite{Mai23,Leb24}.

Several different types of double-heavy hadrons can be described in the B\nobreakdash-O approximation. A bound state in potentials that increase linearly at large distances or a deeply bound state in potentials that are constant at large distances is a compact double-heavy hadron. A loosely bound state in potentials that are constant at large distances is a heavy-hadron molecule. An unbound state in potentials that are constant at large distances is a scattering state of a pair of heavy hadrons. The decay of a double-heavy hadron into a pair of heavy hadrons is mediated by B\nobreakdash-O transition rates between the corresponding potentials, which can in principle be calculated using lattice QCD with two static color sources. The decay width can then be calculated by solving a coupled-channel Schr\"odinger equation where the mixing potential is expressed as a sum of products of angular-momentum coefficients and B\nobreakdash-O transition rates \cite{Bru23a,Braa24}.

The aim of this paper is to derive model-independent  results for the decays of double-heavy hadrons into pairs of heavy hadrons using the techniques introduced in Ref.~\cite{Braa24} for the analogous decays of hidden-heavy hadrons. Constraints from identical particles and from flavor symmetries play a more prominent role in the double-heavy case studied here than in the hidden-heavy case of Ref.~\cite{Braa24}. Heavy quarks have position, spin, and color degrees of freedom. The B\nobreakdash-O wave functions solely depend on the slow degrees of freedom, that is, the positions and spins of the heavy quarks. The heavy-quark colors, on the other hand, are fast degrees of freedom that are hidden within the B\nobreakdash-O potentials. One therefore cannot naively require the B\nobreakdash-O wave functions to be antisymmetric under exchanges of identical heavy quarks. One must instead impose antisymmetry of the double-heavy-hadron state under exchange of identical heavy quarks using both the B\nobreakdash-O potentials and the wave functions.

The rest of this paper is organized as follows. In Sec.~\ref{doubleheavysection}, we review the B\nobreakdash-O expansion for a double-heavy hadron and derive an exclusion principle from identical heavy quarks. In Sec.~\ref{hadronpairsection}, we review the B\nobreakdash-O expansion for a heavy-hadron pair and derive an exclusion principle from identical heavy hadrons. In Sec.~\ref{decaysection}, we write down the coupling potential between a double-heavy hadron and a pair of heavy hadrons. Then, we derive model-independent selection rules and relative partial decay rates for the decays of double-heavy hadrons into pairs of heavy hadrons. In Sec.~\ref{examplesection}, we discuss explicit applications of our general results to conventional double-heavy baryons and to exotic double-heavy tetraquark mesons. Finally, we summarize our conclusions in Sec.~\ref{conclusionsection}.

\section{Double-Heavy Hadrons}
\label{doubleheavysection}

\subsection{Static Limit}
\label{staticlimitsubsection}

QCD is the quantum field theory of the strong interactions between hadrons. It has exact symmetries under the Poincar\'e group, charge conjugation $\mathcal{C}$, and parity transform $\mathcal{P}$. The degrees of freedom of QCD are gluon and quark fields. There are six flavors of quarks in total, with three light flavors $u$, $d$, and $s$ and three heavy flavors $c$, $b$, and $t$. QCD has an approximate $\mathrm{SU}(3)$ symmetry under transformations of the three light flavors $u$, $d$, and $s$. The $\mathrm{SU}(2)$ flavor-symmetry subgroup for the two lightest flavors $u$ and $d$ is a more accurate symmetry, since the strange quark $s$ is significantly heavier. For historical reasons rooted in nuclear physics, the generator of the $\mathrm{SU}(2)$ flavor-symmetry subgroup is referred to as the isospin (isotopic-spin) vector $\bm{I}$.

For the sake of simplicity, in this paper we restrict ourselves to two light flavors with $\mathrm{SU}(2)$ flavor symmetry generated by the isospin vector $\bm{I}$. Our results can be quite easily extended to the case of three light flavors with $\mathrm{SU}(3)$ flavor symmetry or to the physical case of $2+1$ light flavors with approximate $\mathrm{SU}(3)$ and $\mathrm{SU}(2)$ flavor symmetries.

A \emph{double-heavy hadron} is one that contains two heavy quarks $QQ$ or two heavy antiquarks $\bar{Q}\bar{Q}$ with the same flavor $Q\equiv c$ or $b$.%
\footnote{The top flavor $t$ is irrelevant since top (anti)quarks decay weakly before a hadron can be formed.}
We focus explicitly on the $QQ$ case, with the understanding that our results can be easily translated to the $\bar{Q}\bar{Q}$ case using charge conjugation symmetry. A double-heavy hadron also contains gluons $g$ and light quarks $q$ and antiquarks $\bar{q}$ which together with the $QQ$ pair form a color singlet. In the \emph{heavy-quark limit} (heavy-quark mass to infinity), the heavy-quark spins decouple from the gluons and light quarks and antiquarks. Further, in the \emph{static limit} (heavy-quark velocity to zero), the $QQ$ pair has no motion. In this situation, a double-heavy hadron reduces to the system of gluons and light quarks and antiquarks bound to two static color-triplet $(\bm{3})$ sources.

We refer to QCD with only gluon fields and light-quark fields as \emph{light QCD.} Light QCD with two static color sources has the same symmetries as QCD under time translations, time inversion, and flavor transformations. However, the two static color sources break the symmetries under spatial translations, Lorentz boosts, rotations, parity, and charge conjugation. The residual symmetry group consists of the so-called \emph{B\nobreakdash-O symmetries}, also referred to as cylindrical symmetries. For $(\bm{3},\bm{3})$ sources located at $\bigl(+\frac{1}{2}\bm{r},-\frac{1}{2}\bm{r}\bigr)$ with $\bm{r}$ their separation vector, the B\nobreakdash-O symmetries are:
\begin{enumerate}[(i)]
\item Rotations around the axis passing through the two sources, generated by $\bm{J}_\textup{light}\cdot\bm{\hat{r}}$ with $\bm{J}_\textup{light}$ the total angular-momentum vector of the light-QCD fields and $\bm{\hat{r}}$ the unit vector in the direction of $\bm{r}$.
\item Parity transform $\mathcal{P}$ through the origin.
\item Reflections $\mathcal{R}$ through any plane containing the two sources.
\end{enumerate}

The eigenstates of light QCD in the presence of static $(\bm{3},\bm{3})$ sources at $\bigl(+\frac{1}{2}\bm{r},-\frac{1}{2}\bm{r}\bigr)$, or \emph{light-QCD eigenstates} for short, can be labeled using quantum numbers associated with isospin and B\nobreakdash-O symmetries. We denote by $(I,I_3)$ the quantum numbers for $\bm{I}$. We denote by $\lambda$ the quantum number for $\bm{J}_\textup{light}\cdot\bm{\hat{r}}$. We denote by $\eta$ the quantum number for parity $\mathcal{P}$ acting on the light-QCD fields. We denote by $\epsilon$ the quantum number for reflections $\mathcal{R}$, which is defined only for light-QCD states with $\lambda=0$. We refer to the quantum numbers $\lambda$ and $\eta$ (and $\epsilon$ if $\lambda=0$) as \emph{B\nobreakdash-O quantum numbers.} Symmetries under reflections $\mathcal{R}$ imply that states with opposite values of $\lambda$ have the same energy. Therefore, an alternate choice for the B\nobreakdash-O quantum numbers is $\Lambda=\lvert\lambda\rvert$, $\eta$, and $\epsilon$. These quantum numbers are traditionally denoted by $\Lambda_\eta^\epsilon$. The subscript $\eta$ is $g$ or $u$ if the quantum number $\eta$ is $+1$ or $-1$. The superscript $\epsilon$, specified only if $\Lambda=0$, is $+$ or $-$ if the quantum number $\epsilon$ is $+1$ or $-1$. For historical reasons rooted in atomic physics, integer values of $\Lambda$ are denoted by an uppercase Greek letter instead of a number, following the convention $\Lambda\to \Sigma,\Pi,\Delta,\Phi$ for $\Lambda=0,1,2,3$, and so on.

In the limit $\bm{r}\to0$, two static color sources at $\bigl(+\frac{1}{2}\bm{r},-\frac{1}{2}\bm{r}\bigr)$ act as a linear combination of single color sources at the origin. If the sources are $(\bm{3},\bm{3})$, they act as a linear combination of a $\bm{3}^\ast$ source and of a $\bm{6}$ source. Light QCD with a single color source at the origin is symmetric under all rotations generated by $\bm{J}_\textup{light}$ plus parity inversion through the origin. Thus, its eigenstates have well-defined quantum numbers $(j,m)$ for $\bm{J}_\textup{light}$ and $\pi$ for $\mathcal{P}$. We denote the multiplets of $2j+1$ states by $j^\pi$.

Because of the rotational and parity symmetries, the light-QCD eigenstates must form degenerate multiplets in the limit $\bm{r}\to0$. These multiplets have definite light-QCD angular-momentum and parity quantum numbers $j^\pi$. They consist of $2j+1$ eigenstates with the same B\nobreakdash-O quantum number $\eta$ and with $\lambda$ ranging from $-j$ to $+j$. In the short-distance region where $r$ is smaller than the size of a light-QCD state bound to a single color source, the light-QCD eigenstates have approximate quantum numbers $j^\pi$. Beyond this short-distance region, $j^\pi$ cease being good quantum numbers. Instead, each light-QCD eigenstate with B\nobreakdash-O quantum numbers $\Lambda_\eta^\epsilon$ will be expressed as an infinite expansion in light-QCD states with definite quantum numbers $j^\pi$ with $j\geq\Lambda$ and $\pi=-1,+1$. In practical applications of the B\nobreakdash-O approximation, however, one applies a truncation to a finite number of light-QCD eigenstates to obtain a closed system of coupled Schr\"odinger equations. It can be shown that if such a truncation is warranted, then the infinite $j^\pi$ expansion of each light-QCD eigenstate can also be truncated. It is then possible to perform an (approximately) unitary transformation from the \emph{adiabatic} basis of light-QCD eigenstates that are a superposition of various $j^\pi$ to the \emph{diabatic} basis of states that have definite quantum numbers $j^\pi$ \cite{Bru23a}. These so-called \emph{diabatic light-QCD states} are thus labeled by the angular-momentum and parity quantum numbers $j^\pi$ in addition to the B\nobreakdash-O quantum numbers $\lambda$ and $\eta$.

The expectation values of the light-QCD Hamiltonian for states in the diabatic basis define \emph{diabatic B\nobreakdash-O potentials} that are dominated by light-QCD angular momentum and parity $j^\pi$ within the validity region of a given truncation. Since the diabatic states are not eigenstates of the light-QCD Hamiltonian, there can be transitions between diabatic B\nobreakdash-O potentials with the same $\Lambda_\eta^\epsilon$ quantum numbers (regardless of their other quantum numbers $j^\pi$). Transitions between two diabatic B\nobreakdash-O potentials that are both confining at large distances generate mixing effects between double-heavy hadrons. Transitions between a B\nobreakdash-O potential that is confining at large distances and one that is constant at large distances generate both mixing effects and decays of double-heavy hadrons into pairs of heavy hadrons.

In what follows, we shall work exclusively within the diabatic basis. So, for the sake of brevity, we will omit the qualifier ``diabatic'' when referring to the light-QCD states and B\nobreakdash-O potentials.

We denote by a ket $\smallvket{(+,-)}{j^\pi}{\lambda}{,\eta}$ a light-QCD state in the presence of $(\bm{3},\bm{3})$ sources at $\bigl(+\frac{1}{2}\bm{r},-\frac{1}{2}\bm{r}\bigr)$, with B\nobreakdash-O quantum numbers $\lambda$ and $\eta$ and light-QCD angular-momentum and parity $j^\pi$. We denote by a ket $\smallvket{}{I}{I_3}{}$ its isospin state with quantum numbers $(I,I_3)$.
The light-QCD state is completely specified by the direct product
\begin{equation}
\vket{(+,-)}{j^\pi}{\lambda}{,\eta} \vket{}{I}{I_3}{}.
\label{lightqcdstate}
\end{equation}
Note that the light-QCD state in Eq.~\eqref{lightqcdstate} depends parametrically on the coordinates $\bigl(+\frac{1}{2}\bm{r},-\frac{1}{2}\bm{r}\bigr)$ of the static color sources. We indicate this parametric dependence through the arguments $(+,-)$ inside $\smallvket{(+,-)}{j^\pi}{\lambda}{,\eta}$. Transformations acting on the light-QCD state may also affect the coordinates of the static sources, as explained next.

Now consider the action of parity $\mathcal{P}$ on the light-QCD state in Eq.~\eqref{lightqcdstate}. Since it is an eigenstate of $\mathcal{P}$ with eigenvalue $\eta$, its action is
\begin{equation}
\mathcal{P}\vket{(+,-)}{j^\pi}{\lambda}{,\eta}  = \eta \vket{(+,-)}{j^\pi}{\lambda}{,\eta}.
\label{parity}
\end{equation}
Alternatively, the action of $\mathcal{P}$ on the light-QCD state in Eq.~\eqref{lightqcdstate} amounts to the inversion $\bm{r}\to-\bm{r}$, that is, exchanging the positions of the arguments $+$ and $-$, and multiplying by the parity $\pi$:
\begin{equation}
\mathcal{P}\vket{(+,-)}{j^\pi}{\lambda}{,\eta} = \pi \vket{(-,+)}{j^\pi}{\lambda}{,\eta}.
\label{parityrzero}
\end{equation}
Note that the B\nobreakdash-O quantum number $\eta$ of a double-source light-QCD state is generally distinct from the parity $\pi$ of the single-source light-QCD state associated to its $\bm{r}\to0$ multiplet. Comparing the right sides of Eqs.~\eqref{parity} and \eqref{parityrzero}, we see that their product $\pi\eta$ is determined by the symmetry of the ket $\smallvket{(+,-)}{j^\pi}{\lambda}{,\eta}$ under the inversion $\bm{r}\to-\bm{r}$:
\begin{equation}
\vket{(-,+)}{j^\pi}{\lambda}{,\eta} = \pi\, \eta\, \vket{(+,-)}{j^\pi}{\lambda}{,\eta}.
\label{rminusr}
\end{equation}

The two heavy quarks in the light-QCD state of a double-heavy hadron can be treated as static color sources. So, the exchange of the source positions $\bm{r}\to-\bm{r}$ in Eq.~\eqref{rminusr} can be reduced to the exchange of the color indices of the static sources. This observation leads to a direct physical interpretation of the relative sign between $\eta$ and $\pi$. Let us first realize that the product $\pi\eta$ can be either $+1$ or $-1$ for any value of $\bm{r}$, so continuity in $\bm{r}$ requires that $\pi\eta$ is a constant for all $\bm{r}$. We can then calculate $\pi\eta$ in Eq.~\eqref{rminusr} in the limit $\bm{r}\to0$ and the result will be valid in general. In the case of $(\bm{3},\bm{3})$ sources acting as a $\bm{3}^\ast$ source, which correspond to B-O potentials that are \emph{attractive} at short range, the light-QCD states are antisymmetric under color exchange and therefore $\pi\eta=-1$. In the case of $(\bm{3},\bm{3})$ sources acting as a $\bm{6}$ source, which correspond to B-O potentials that are \emph{repulsive} at short range, the light-QCD states are symmetric under color exchange and therefore $\pi\eta=+1$. We thus obtain a simple rule connecting the relative sign between $\pi$ and $\eta$ to the behavior of the corresponding B\nobreakdash-O potentials at short range:
\begin{equation}
\pi \eta =
\begin{cases}
-1 & \text{for attractive potentials,} \\
+1 & \text{for repulsive potentials.} \\
\end{cases}
\label{colorsym}
\end{equation}
It is important to emphasize that our convention is to take the fermionic nature of the heavy quark into account in its spin state rather than in the static color source. This convention has been used previously by Ref.~\cite{Bic16}. The alternative convention is to take the fermionic nature into account in the static color source, in which case the source must be treated as a Grassman number. This convention has been used previously by Refs.~\cite{Naj09,Sot20b}. If adopting this alternative convention, one should substitute $\eta\to-\eta$.

Finally, let us note that the light-QCD state in Eq.~\eqref{lightqcdstate} has definite transformations under reflections $\mathcal{R}$ through a plane containing the static sources. Without loss of generality, let $\bm{\hat{z}}$ be the axis passing through the sources and $zx$ be the reflection plane. The operator $\mathcal{R}$ can then be expressed as a parity transform $\mathcal{P}$ followed by the rotation of an angle $\pi$ around the axis $\bm{\hat{y}}$,
\begin{equation}
\mathcal{R} = e^{-i\pi \bm{J}\cdot\bm{\hat{y}}} \mathcal{P}.
\end{equation}
Using Eq.~\eqref{parityrzero} and
\begin{equation}
e^{-i\pi \bm{J}\cdot\bm{\hat{y}}} \vket{(-,+)}{j^\pi}{\lambda}{,\eta} = (-1)^{j-\lambda} \vket{(+,-)}{j^\pi}{-\lambda}{,\eta},
\end{equation}
we have
\begin{equation}
\mathcal{R} \vket{(+,-)}{j^\pi}{\lambda}{,\eta} = \pi (-1)^{j-\lambda} \vket{(+,-)}{j^\pi}{-\lambda}{,\eta}.
\label{reflectioneq}
\end{equation}
From Eq.~\eqref{reflectioneq}, we see that the light-QCD states in Eq.~\eqref{lightqcdstate} with $\lambda=0$ are eigenstates of reflections $\mathcal{R}$ with quantum number
\begin{equation}
\epsilon = \pi (-1)^j.
\label{lightqcdreflection}
\end{equation}

\subsection{Orbital Angular Momentum and Spin}

The spin and parity quantum numbers of a heavy quark $Q$ are $\frac{1}{2}^+$. We denote by a ket $\smallvket{}{\frac{1}{2}^+}{m}{(\pm)}$ the two spin and parity states of a heavy quark at $\pm\frac{1}{2}\bm{r}$ with $m=-\frac{1}{2},+\frac{1}{2}$. The four independent spin states of a $QQ$ pair at $\bigl(+\frac{1}{2}\bm{r},-\frac{1}{2}\bm{r}\bigr)$ can be decomposed into states with total heavy-spin quantum numbers $(S_Q,m_{S_Q})$:
\begin{equation}
\vket{\Bigl(\tfrac{1}{2}^+(+), \tfrac{1}{2}^+(-)\Bigr)}{S_Q}{m_{S_Q}}{}
= \sum_{m, m^\prime} \cg{\tfrac{1}{2}}{m}{\tfrac{1}{2}}{m^\prime}{S_Q}{m_{S_Q}} \vket{}{\tfrac{1}{2}^+}{m}{(+)} \vket{}{\tfrac{1}{2}^+}{m^\prime}{(-)},
\label{spinstate}
\end{equation}
where $\smallcg{j_1}{m_1}{j_2}{m_2}{j_3}{m_3}$ is a Clebsch-Gordan coefficient. The heavy-quark-spin states in Eq.~\eqref{spinstate} have definite symmetry under the inversion $\bm{r}\to-\bm{r}$:
\begin{equation}
\vket{\Bigl(\tfrac{1}{2}^+(-), \tfrac{1}{2}^+(+)\Bigr)}{S_Q}{m_{S_Q}}{}
= (-1)^{S_Q} \vket{\Bigl(\tfrac{1}{2}^+(+), \tfrac{1}{2}^+(-)\Bigr)}{S_Q}{m_{S_Q}}{},
\label{qqrminusr}
\end{equation}
where the sign $(-1)^{S_Q}$ on the right side is the product of a factor $(-1)^{S_Q - 1}$ from the symmetries of Clebsch-Gordan coefficients and a factor $-1$ from changing the order of the fermionic operators for the two heavy quarks.%
\footnote{This last factor $-1$ is necessary in our convention, because we take into account the fermionic nature of the heavy quark into its spin state. In the alternative convention where the fermionic nature is taken into account in the static color source, the sign factor in Eq.~\eqref{qqrminusr} is $(-1)^{S_Q-1}$.}
Applying parity $\mathcal{P}$ to both sides of Eq.~\eqref{spinstate} yields
\begin{equation}
\mathcal{P}\vket{\Bigl(\tfrac{1}{2}^+(+), \tfrac{1}{2}^+(-)\Bigr)}{S_Q}{m_{S_Q}}{} = \vket{\Bigl(\tfrac{1}{2}^+(-), \tfrac{1}{2}^+(+)\Bigr)}{S_Q}{m_{S_Q}}{}.
\label{qqparity}
\end{equation}
The positions of the arguments $-$ and $+$ on the right side can be exchanged using Eq.~\eqref{qqrminusr}.

In the B\nobreakdash-O approximation, a double-heavy-hadron state is expressed as an integral over $\bm{r}$ of the $\bm{r}$-dependent kets for the light-QCD and heavy-spin states multiplied by wave functions. The B\nobreakdash-O wave functions that depend on the vector $\bm{r}$ can be expanded in terms of radial wave functions that depend on the distance $r = \lvert \bm{r} \rvert$ and spherical harmonics $Y_{L_Q}^{m_{L_Q}}(\bm{\hat{r}})$ that depend on the unit vector $\bm{\hat{r}}$, where $(L_Q, m_{L_Q})$ are the orbital-angular-momentum quantum numbers of the $QQ$ pair. We define the \emph{B\nobreakdash-O angular momentum} vector $\bm{L}$ as the sum of the light-QCD angular momentum  $\bm{J}_\textup{light}$ and the $QQ$ orbital angular momentum $\bm{L}_Q$:
\begin{equation}
\bm{L} = \bm{J}_\textup{light} + \bm{L}_Q.
\label{bomomentum}
\end{equation}
The B\nobreakdash-O angular momentum is useful, because it is conserved in the heavy-quark limit. We denote its associated quantum numbers by $(L,M_L)$. The sum of the B\nobreakdash-O angular momentum $\bm{L}$ and the total heavy spin $\bm{S}_Q$ gives the total angular momentum $\bm{J}$:
\begin{equation}
\bm{J} = \bm{L} + \bm{S}_Q,
\end{equation}
which is exactly conserved. We denote its quantum numbers by $(J,M_J)$. The spins of the heavy quarks are conserved up to corrections suppressed by negative powers of the heavy-quark mass. Thus, double-heavy hadrons with the same B\nobreakdash-O angular momentum $L$ and parity $P$ but different total angular momentum $J$ form approximately degenerate \emph{B\nobreakdash-O multiplets}. The multiplets are labeled by the  B\nobreakdash-O angular-momentum and parity $L^P$ and other quantum numbers.

The state of a double-heavy hadron in B\nobreakdash-O potentials dominated by light-QCD angular-momentum and parity $j^\pi$ is labeled by the total angular momentum and parity quantum numbers $(J^P, M_J)$, the isospin quantum numbers $(I, I_3)$, the total heavy spin $S_Q$, the B\nobreakdash-O angular momentum $L$, the light-QCD parity $\eta$, and other quantum numbers. The B\nobreakdash-O expansion of a double-heavy-hadron state reads
\begin{align}
\Bigl\lvert\Psi_{J^P,I,S_Q,L,\eta,j^\pi}^{M_J,I_3}\Bigr\rangle =& \vket{}{I}{I_3}{} \sum_{L_Q} \int \mathrm{d}^3 \bm{r} \, \Psi_{L_Q,J^P,I,S_Q,L,\eta,j^\pi}(r) \notag \\
&\times\sum_{m_{S_Q},M_L} \cg{S_Q}{m_{S_Q}}{L}{M_L}{J}{M_J} \vket{\Bigl(\tfrac{1}{2}^+(+), \tfrac{1}{2}^+(-)\Bigr)}{S_Q}{m_{S_Q}}{} \notag\\
&\times\sum_{\lambda,m_{L_Q}} \cg{j}{\lambda}{L_Q}{m_{L_Q}}{L}{M_L} \vket{(+,-)}{j^\pi}{\lambda}{,\eta} Y_{L_Q}^{m_{L_Q}}(\bm{\hat{r}}),
\label{doubleheavyexpansion}
\end{align}
where we have moved the ket $\smallvket{}{I}{I_3}{}$ outside of the integral since it does not depend on $\bm{r}$. Note that Eq.~\eqref{doubleheavyexpansion} contains a sum over the $QQ$ orbital angular momentum $L_Q$ because it is not conserved in general. The parity transform of the spherical harmonics is
\begin{equation}
Y_{L_Q}^{m_{L_Q}}(-\bm{\hat{r}}) = (-1)^{L_Q} Y_{L_Q}^{m_{L_Q}}(\bm{\hat{r}}).
\label{orbitalrminusr}
\end{equation}
Applying parity $\mathcal{P}$ to both sides of Eq.~\eqref{doubleheavyexpansion} and then using Eqs.~\eqref{parity}, \eqref{qqparity} and \eqref{orbitalrminusr}, we see that the sum over $L_Q$ in Eq.~\eqref{doubleheavyexpansion} is constrained to either even or odd values, according to
\begin{equation}
P = \pi (-1)^{L_Q}.
\label{exactparity}
\end{equation}
It is safe to assume that the lowest-energy double-heavy-hadron states contain the smallest orbital angular momentum $L_Q=0$ in their B\nobreakdash-O expansion. Under this assumption, Eqs.~\eqref{bomomentum} and \eqref{exactparity} require that the B\nobreakdash-O angular momentum and parity of the lowest double-heavy hadron multiplet are
\begin{equation}
L^P=j^\pi.
\label{lowestbomultiplet}
\end{equation}

Equations~\eqref{rminusr}, \eqref{qqrminusr}, and \eqref{orbitalrminusr}, indicate that each $\bm{r}$-dependent factor inside the integral on the right side of Eq.~\eqref{doubleheavyexpansion} has a definite symmetry under the inversion $\bm{r}\to-\bm{r}$. Because of this, the integral in Eq.~\eqref{doubleheavyexpansion} is automatically zero unless
\begin{equation}
\pi \eta (-1)^{S_Q+L_Q} = +1.
\label{rawexclusionprinciple}
\end{equation}
Equation~\eqref{rawexclusionprinciple} is written in terms of quantum numbers $\eta$ and $S_Q$ that are conserved in the heavy-quark limit and quantum numbers $\pi$ and $L_Q$ that are generally not conserved. The dependence of Eq.~\eqref{rawexclusionprinciple} on the nonconserved quantum numbers $\pi$ and $L_Q$ can be absorbed into the exactly-conserved quantum number $P$ using Eq.~\eqref{exactparity}, which gives
\begin{equation}
P = \eta (-1)^{S_Q}.
\label{boexclusionprinciple}
\end{equation}
Equation~\eqref{boexclusionprinciple} is the \emph{B\nobreakdash-O exclusion principle} for double-heavy hadrons. It states that the total heavy spin of a double-heavy hadron in a B\nobreakdash-O potential $\Lambda_\eta^\epsilon$ is necessarily $S_Q=0$ if its parity is $P=\eta$ and $S_Q=1$ if its parity is $P=-\eta$.

Interestingly, the B\nobreakdash-O exclusion principle in Eq.~\eqref{boexclusionprinciple} mirrors the well-known exclusion principle from identical quarks in quark models. To see this, we multiply both sides of Eq.~\eqref{rawexclusionprinciple} by $-1$ and factor the left side into terms for color, spin, and position exchange of identical heavy quarks, which yields the expected expression:
\begin{equation}
\underbrace{\vphantom{(}\pi\eta}_\text{color} \times \underbrace{(-1)^{S_Q+1}}_\text{spin} \times \underbrace{(-1)^{L_Q}}_\text{position} = -1.
\end{equation}

\section{Hadron Pairs}
\label{hadronpairsection}

\subsection{Single Heavy Hadron}

A \emph{heavy hadron} is one that contains a heavy quark $Q$ or a heavy antiquark $\bar{Q}$. A heavy hadron also contains light-QCD fields which together with the $Q$ or $\bar{Q}$ form a color singlet. In the heavy-quark limit (heavy-quark mass to infinity), the spin of the $Q$ or $\bar{Q}$ decouples from the light-QCD fields. Further, in the static limit (heavy-quark velocity to zero), the $Q$ or $\bar{Q}$ has no motion. In this situation, a heavy hadron reduces to the system of light-QCD fields bound to a static $\bm{3}$ or $\bm{3}^\ast$ color source. We refer to the light-QCD fields together with the static color source as a \emph{static hadron}. It is a \emph{static meson} if it has the flavor of a light quark $q$ or antiquark $\bar{q}$. It is a \emph{static baryon} or \emph{static antibaryon} if it has the flavor of two light quarks $qq$ or two light antiquarks $\bar{q}\bar{q}$, respectively.

A static hadron at the position $\pm\frac{1}{2}\bm{r}$ can be labeled by angular-momentum quantum numbers $(j,m)$ for rotations around any axis passing through the point $\pm\frac{1}{2}\bm{r}$. It also has a quantum number $\pi$ for the parity transform through the point $\pm\frac{1}{2}\bm{r}$. We denote these angular-momentum and parity quantum numbers of a static hadron by $j^\pi(\pm)$, where the argument $\pm$ indicates that they apply to the light-QCD fields generated by the source at $\pm\frac{1}{2}\bm{r}$. A static hadron at a position $\bm{r}$ can be also labeled by isospin quantum numbers $(i,\tau)$. We denote the isospin quantum number of a static hadron by $i(\pm)$, where the argument $\pm$ indicates that they apply to the light-QCD fields generated by the source at $\pm\frac{1}{2}\bm{r}$.

We denote by a ket $\smallvket{}{j^\pi}{m}{(\pm)}$ the $2j+1$ angular-momentum and parity states of a static hadron at $\pm\frac{1}{2}\bm{r}$ with $m=-j,\dotsc,+j$. We denote its $2 i + 1$ isospin states by a ket $\smallvket{}{i}{\tau}{(\pm)}$. The state of a heavy hadron that contains a heavy quark $Q$ at $\pm\frac{1}{2}\bm{r}$ can be constructed from the direct product of the angular-momentum and parity state of a static hadron and the spin and parity state of a heavy quark. The heavy-hadron state with light-QCD angular-momentum and parity $j^\pi$ and with spin $J= \bigl\lvert j - \frac{1}{2}\bigr\rvert$ or $j + \frac{1}{2}$ is
\begin{equation}
\vket{(\tfrac{1}{2}^{+}, j^{\pi} )}{J}{M}{(\pm)} = \sum_{m, m^\prime} \cg{\tfrac{1}{2}}{m}{j}{m^\prime}{J}{M} \vket{}{\tfrac{1}{2}^+}{m}{(\pm)} \vket{}{j^{\pi}}{m^\prime}{(\pm)},
\label{hadef}
\end{equation}
where $M=-J,\dotsc,+J$. The spin of the heavy quark is conserved up to corrections suppressed by negative powers of the heavy-quark mass. Thus, heavy hadrons corresponding to the same static hadron $j^\pi$ but with different spins $J$ form an approximately degenerate multiplet of \emph{Heavy-Quark Spin Symmetry} (HQSS). The HQSS multiplets are labeled by the angular-momentum parity $j^\pi$ and other quantum numbers.

\subsection{Pair of Static Hadrons}

Let us consider a pair of static hadrons at $\bigl(+\frac{1}{2}\bm{r},-\frac{1}{2}\bm{r}\bigr)$ with angular-momentum and parity $(j_1^{\pi_1}(+),j_2^{\pi_2}(-))$ and with isospin $(i_1(+),i_2(-))$, respectively. The $(2 j_1 + 1) (2 j_2 + 1)$ angular-momentum and parity states of the pair can be decomposed into states with light-QCD angular-momentum quantum numbers $(j^\prime,\lambda)$:
\begin{equation}
\vket{\Bigl(j_1^{\pi_1} (+), j_2^{\pi_2} (-) \Bigr)}{j^\prime}{\lambda}{} \\
= \sum_{m, m^\prime} \cg{j_1}{m}{j_2}{m^\prime}{j^\prime}{\lambda} \vket{}{j_1^{\pi_1}}{m}{(+)} \vket{}{j_2^{\pi_2}}{m^\prime}{(-)}.
\label{pairangularparity}
\end{equation}
The $(2 i_1 + 1) (2 i_2 + 1)$ isospin states of the pair can be decomposed into states with total isospin quantum numbers $(I,I_3)$:
\begin{equation}
\vket{\Bigl(i_1 (+), i_2 (-) \Bigr)}{I}{I_3}{} = \sum_{\tau, \tau^\prime} \cg{i_1}{\tau}{i_2}{\tau^\prime}{I}{I_3} \vket{}{i_1}{\tau}{(+)} \vket{}{i_2}{\tau^\prime}{(-)}.
\label{pairisospin}
\end{equation}
The light-QCD state of the hadron-pair is the direct product of the angular-momentum and parity state in Eq.~\eqref{pairangularparity} and the isospin state in Eq.~\eqref{pairisospin}:
\begin{equation}
\vket{\Bigl(j_1^{\pi_1} (+), j_2^{\pi_2} (-) \Bigr)}{j^\prime}{\lambda}{}\vket{\Bigl(i_1 (+), i_2 (-) \Bigr)}{I}{I_3}{}.
\label{staticpair}
\end{equation}

In general, the static-hadron-pair states in Eq.~\eqref{staticpair} are not eigenstates of parity inversion $\mathcal{P}$ through the origin. Their transformation property under $\mathcal{P}$ is
\begin{equation}
\mathcal{P}\vket{\Bigl(j_1^{\pi_1} (+), j_2^{\pi_2} (-) \Bigr)}{j^\prime}{\lambda}{} \vket{\Bigl(i_1 (+), i_2 (-) \Bigr)}{I}{I_3}{}
= \pi_1 \pi_2\vket{\Bigl(j_1^{\pi_1} (-), j_2^{\pi_2} (+) \Bigr)}{j^\prime}{\lambda}{} \vket{\Bigl(i_1 (-), i_2 (+) \Bigr)}{I}{I_3}{},
\label{staticpairparity}
\end{equation}
where we have used the fact that inverting $\bm{r}\to-\bm{r}$ amounts to exchanging the positions of the arguments $+$ and $-$.
The state on the right side is independent from that on the left side unless the two static hadrons are identical. For a pair of identical static hadrons, which requires $j_2^{\pi_2}=j_1^{\pi_1}$ and $i_2 = i_1$, the state in Eq.~\eqref{staticpair} has a definite symmetry under the inversion $\bm{r}\to-\bm{r}$:
\begin{multline}
\vket{\Bigl(j_1^{\pi_1} (-), j_1^{\pi_1} (+) \Bigr)}{j^\prime}{\lambda}{} \vket{\Bigl(i_1 (-), i_1 (+) \Bigr)}{I}{I_3}{} \\
= (-1)^{j^\prime + I - 2 i_1} \vket{\Bigl(j_1^{\pi_1} (+), j_1^{\pi_1} (-) \Bigr)}{j^\prime}{\lambda}{} \vket{\Bigl(i_1 (+), i_1 (-) \Bigr)}{I}{I_3}{},
\label{staticpairidenticalrminusr}
\end{multline}
where the sign is the product of the factors $(-1)^{j^\prime - 2 j_1}$ and $(-1)^{I - 2 i_1}$ from the symmetries of the Clebsch-Gordan coefficients and a factor $(-1)^{2 j_1}$ from changing the order of the operators for the two static hadrons. Therefore, the state of a pair of identical static hadrons is a simultaneous eigenstate of parity with eigenvalue
\begin{equation}
\eta = (-1)^{j^\prime + I - 2 i_1},
\label{staticpairidenticaleta}
\end{equation}
as can be checked by plugging Eq.~\eqref{staticpairidenticalrminusr} into Eq.~\eqref{staticpairparity} with $j_2^{\pi_2}=j_1^{\pi_2}$ and $i_2 = i_1$. If the two static hadrons are not identical, one can construct static-hadron-pair states with either $\eta=+1$ or $-1$ by applying the parity projector 
\begin{equation}
\Pi_\eta = \frac{\mathbbm{1} + \eta \mathcal{P}}{2}
\end{equation}
onto the states in Eq.~\eqref{staticpair}:
\begin{equation}
\Pi_\eta \vket{\Bigl(j_1^{\pi_1} (+), j_2^{\pi_2} (-) \Bigr)}{j^\prime}{\lambda}{}\vket{\Bigl(i_1 (+), i_2 (-) \Bigr)}{I}{I_3}{}.
\label{staticpairproject}
\end{equation}
Note that the $\eta$-projected static-hadron-pair states in Eq.~\eqref{staticpairproject}, unlike the nonprojected states in Eq.~\eqref{staticpair}, do have a definite symmetry under the inversion $\bm{r}\to-\bm{r}$:
\begin{multline}
\Pi_\eta\vket{\Bigl(j_1^{\pi_1} (-), j_2^{\pi_2} (+) \Bigr)}{j^\prime}{\lambda}{} \vket{\Bigl(i_1 (-), i_2 (+) \Bigr)}{I}{I_3}{} \\
= \pi_1 \pi_2 \eta \, \Pi_\eta \vket{\Bigl(j_1^{\pi_1} (+), j_2^{\pi_2} (-) \Bigr)}{j^\prime}{\lambda}{} \vket{\Bigl(i_1 (+), i_2 (-) \Bigr)}{I}{I_3}{}.
\label{staticpairprojectrminusr}
\end{multline}

The $\bm{r}$-dependent kets in Eq.~\eqref{staticpairproject} define hadron-pair potentials with B\nobreakdash-O quantum numbers $\Lambda_\eta^\epsilon$ where $\Lambda\leq j^\prime$. If the two static hadrons are identical, then the value of $\eta$ is constrained by $j^\prime$ and $I$ according to Eq.~\eqref{staticpairidenticaleta}. Otherwise, the static-hadron pair defines B\nobreakdash-O potentials with $\eta=+1$ and $\eta=-1$ for any value of $j^\prime$ and $I$. For $\Lambda=0$, the reflection quantum number $\epsilon$ is
\begin{equation}
\epsilon=\pi_1 \pi_2 (-1)^{j^\prime}.
\label{pairreflection}
\end{equation}

When the two static hadrons are well separated, a hadron-pair potential is equal to the static-hadron-pair threshold, that is, the sum of the energies of the two static hadrons. When the two static hadrons are close to each other, interactions between the static hadrons cause the hadron-pair-potential to deviate from the static-hadron-pair threshold. Ultimately, as the separation goes to zero, gluonic interactions between the two static color sources dominate and the hadron-pair potential must approach a color-Coulomb potential. As discussed in Sec.~\ref{doubleheavysection}, the color-Coulomb potential potential is either attractive or repulsive depending on the symmetry of the corresponding light-QCD state under the inversion $\bm{r}\to-\bm{r}$. Using Eqs.~\eqref{staticpairidenticalrminusr} and \eqref{staticpairidenticaleta} for identical static hadrons or Eq.~\eqref{staticpairprojectrminusr} for distinct static hadrons, we obtain a simple and general rule:
\begin{equation}
\pi_1 \pi_2 \eta =
\begin{cases}
-1 & \text{for attractive potentials,} \\
+1 & \text{for repulsive potentials.} \\
\end{cases}
\label{hadronsym}
\end{equation}
For a pair of static hadrons with identical light-QCD parity, $\pi_2=\pi_1$, $\Lambda_g^\epsilon$ potentials are repulsive and $\Lambda_u^\epsilon$ potentials are attractive. For a pair of static hadrons with opposite light-QCD parity, $\pi_2=-\pi_1$, $\Lambda_g^\epsilon$ potentials are attractive and $\Lambda_u^\epsilon$ potentials are repulsive. This rule was first deduced in Ref.~\cite{Bic16} from the analysis of 36 independent potentials between pairs of $S$- or $P^-$-wave mesons calculated using lattice QCD. It generalizes what has been found in earlier lattice-QCD calculations of the potentials between pairs of $S$-wave mesons \cite{Mic99,Coo02}.

\subsection{Pair of Heavy Hadrons}

Now consider the state of a pair of heavy hadrons at $\bigl(+\frac{1}{2}\bm{r},-\frac{1}{2}\bm{r}\bigr)$. Let $J_1$ be the spin of the heavy hadron at $+\frac{1}{2}\bm{r}$, $i_1$ its isospin, and $j_1^{\pi_1}$ the angular-momentum and parity of the corresponding static hadron. Let $J_2$ be the spin of the heavy hadron at $-\frac{1}{2}\bm{r}$, $i_2$ its isospin, and $j_2^{\pi_2}$ the angular-momentum and parity of the corresponding static hadron. The states of the heavy-hadron pair are constructed from the direct product of the individual states of each heavy hadron. The isospin state of the pair with total isospin quantum numbers $(I,I_3)$ is defined in Eq.~\eqref{pairisospin}. For its spin and light-QCD angular-momentum states, it is useful to define the \emph{static angular momentum} vector,
\begin{equation}
\bm{J}_\textup{static} = \bm{S}_Q + \bm{J}_\textup{light},
\end{equation}
which generates the rotations of the light-QCD fields and heavy-quark spins. For a pair of heavy hadrons, $\bm{J}_\textup{static}$ is just the total spin of the two heavy hadrons, $\bm{J}_\textup{static} = \bm{J}_1 + \bm{J}_2$. Hence, we denote its quantum numbers by $(S,M_S)$. The heavy-hadron-pair state with static-angular-momentum quantum numbers $(S,M_S)$ is
\begin{multline}
\vket{\Bigl[(\tfrac{1}{2}^+, j_1^{\pi_1}  ) J_1(+), (\tfrac{1}{2}^+, j_2^{\pi_2}) J_2(-) \Bigr]}{S}{M_S}{} \\
= \sum_{M_1,M_2} \cg{J_1}{M_1}{J_2}{M_2}{S}{M_S} \vket{(\tfrac{1}{2}^+, j_1^{\pi_1}  )}{J_1}{M_1}{(+)} \vket{(\tfrac{1}{2}^+, j_2^{\pi_2}  )}{J_2}{M_2}{(-)}.
\label{stateHH}
\end{multline}
The state of the heavy-hadron pair at $\bigl(+\frac{1}{2}\bm{r},-\frac{1}{2}\bm{r}\bigr)$ is the direct product of the static-angular-momentum state in Eq.~\eqref{stateHH} and the isospin state in Eq.~\eqref{pairisospin}:
\begin{equation}
\vket{\Bigl[(\tfrac{1}{2}^+, j_1^{\pi_1}  ) J_1(+), (\tfrac{1}{2}^+, j_2^{\pi_2}) J_2(-) \Bigr]}{S}{M_S}{} \\
\times \vket{\Bigl(i_1 (+), i_2 (-) \Bigr)}{I}{I_3}{}.
\label{stateHHcomplete}
\end{equation}

In the B\nobreakdash-O approximation, a heavy-hadron-pair state is expressed as an integral over $\bm{r}$ of the $\bm{r}$-dependent kets in Eq.~\eqref{stateHHcomplete} multiplied by wave functions. The B\nobreakdash-O wave functions that depend on the vector $\bm{r}$ can be expanded in terms of radial wave functions that depend on the distance $r = \lvert \bm{r} \rvert$ and spherical harmonics $Y_{L_Q^\prime}^{m_{L_Q^\prime}}(\bm{\hat{r}})$ that depend on the unit vector $\bm{\hat{r}}$, where $(L_Q^\prime, m_{L_Q^\prime})$ are the orbital-angular-momentum quantum numbers of the heavy-hadron pair. The sum of the static angular momentum $\bm{J}_\textup{static}$ and the orbital angular momentum $\bm{L}_Q$ is the conserved total angular momentum
\begin{equation}
\bm{J} = \bm{J}_\textup{static} + \bm{L}_Q.
\end{equation}
Its quantum numbers are $(J,M_J)$.

The state of a heavy-hadron pair is labeled by the total angular-momentum and parity quantum numbers $(J^P, M_J)$, the total isospin quantum numbers $(I,I_3)$, the orbital angular momentum $L_Q^\prime$, the static angular momentum $S$, and other quantum numbers. The B\nobreakdash-O expansion of a heavy-hadron-pair state reads
\begin{multline}
\Bigl\lvert \Phi_{J^P,I,L_Q^\prime,S,J_1,J_2,j_1^{\pi_1},j_2^{\pi_2}}^{M_J,I_3} \Bigr\rangle = \int \mathrm{d}^3 \bm{r} \, \Phi_{J^P,I,L_Q^\prime,S,J_1,J_2,j_1^{\pi_1},j_2^{\pi_2}}(r) \vket{\Bigl(i_1 (+), i_2 (-) \Bigr)}{I}{I_3}{} \\
\times \sum_{m_{S},m_{L_Q^\prime}} \cg{S}{M_S}{L_Q^\prime}{m_{L_Q^\prime}}{J}{M_J} \vket{\Bigl[(\tfrac{1}{2}^+, j_1^{\pi_1}  ) J_1(+), (\tfrac{1}{2}^+, j_2^{\pi_2}) J_2(-) \Bigr]}{S}{M_S}{} Y_{L_Q^\prime}^{m_{L_Q^\prime}}(\bm{\hat{r}}).
\label{pairexpansion}
\end{multline}
The parity transform of the $\bm{r}$-dependent kets in Eq.~\eqref{stateHHcomplete} is
\begin{multline}
\mathcal{P} \vket{\Bigl[(\tfrac{1}{2}^+, j_1^{\pi_1}  ) J_1(+), (\tfrac{1}{2}^+, j_2^{\pi_2}) J_2(-) \Bigr]}{S}{M_S}{} \vket{\Bigl(i_1 (+), i_2 (-) \Bigr)}{I}{I_3}{} \\
= \pi_1 \pi_2 \vket{\Bigl[(\tfrac{1}{2}^+, j_1^{\pi_1}  ) J_1(-), (\tfrac{1}{2}^+, j_2^{\pi_2}) J_2(+) \Bigr]}{S}{M_S}{} \vket{\Bigl(i_1 (-), i_2 (+) \Bigr)}{I}{I_3}{}.
\label{pairrminusr}
\end{multline}
Applying parity $\mathcal{P}$ to both sides of Eq.~\eqref{pairexpansion} and then using Eqs.~\eqref{orbitalrminusr} and \eqref{pairrminusr}, we see that the parity of the heavy-hadron-pair state is
\begin{equation}
P = \pi_1 \pi_2 (-1)^{L_Q^\prime}.
\end{equation}

For a pair of identical heavy hadrons, which requires $j_2^{\pi_2}=j_1^{\pi_1}$, $J_2=J_1$, and $i_2 = i_1$, the state in Eq.~\eqref{stateHHcomplete} has a definite symmetry under the inversion $\bm{r}\to-\bm{r}$:
\begin{multline}
\vket{\Bigl[(\tfrac{1}{2}^+, j_1^{\pi_1}  ) J_1(-), (\tfrac{1}{2}^+, j_1^{\pi_1}) J_1(+) \Bigr]}{S}{M_S}{} \vket{\Bigl(i_1 (-), i_1 (+) \Bigr)}{I}{I_3}{} \\
= (-1)^{S + I - 2 i_1} \vket{\Bigl[(\tfrac{1}{2}^+, j_1^{\pi_1}  ) J_1(+), (\tfrac{1}{2}^+, j_1^{\pi_1}) J_1(-) \Bigr]}{S}{M_S}{}\vket{\Bigl(i_1 (+), i_1 (-) \Bigr)}{I}{I_3}{},
\end{multline}
where the sign is the product of the factors $(-1)^{S - 2 J_1}$ and $(-1)^{I - 2 i_1}$ from the symmetries of Clebsch-Gordan coefficients and a factor $(-1)^{2 J_1}$ from changing the order of the operators for the two heavy hadrons. In this case, we see from Eqs.~\eqref{pairrminusr} and \eqref{orbitalrminusr} that each $\bm{r}$-dependent factor inside the integral on the right side of Eq.~\eqref{pairexpansion} has a definite symmetry under the inversion $\bm{r}\to-\bm{r}$. Because of this, the integral in Eq.~\eqref{pairexpansion} is automatically zero unless
\begin{equation}
(-1)^{L_Q^\prime + S + I - 2 i_1} = +1.
\label{pairexclusionprinciple}
\end{equation}
Equation~\eqref{pairexclusionprinciple} is the expected exclusion principle from identical heavy hadrons. To see this, we multiply both sides of Eq.~\eqref{pairexclusionprinciple} by $(-1)^{-2 J_1}$ and factor the left side into terms for flavor, spin, and position exchange:
\begin{equation}
\underbrace{(-1)^{I - 2 i_1}}_\text{flavor} \times \underbrace{(-1)^{S - 2 J_1}}_\text{spin} \times  \underbrace{(-1)^{L_Q^\prime}}_\text{position} = (-1)^{-2 J_1}.
\end{equation}
The sign $(-1)^{-2 J_1}$ on the right is $-1$ for identical baryons with half-integer $J_1$ and $+1$ for identical mesons with integer $J_1$.

\section{Decays of a Double-Heavy Hadron}
\label{decaysection}

\subsection{Coupling Potentials}

A light-QCD state of a double-heavy hadron $j^\pi$ can have a transition into that of a pair of static hadrons $(j_1^{\pi_1}, j_2^{\pi_2}) j^\prime$ with the same B\nobreakdash-O quantum numbers $\lambda$ and $\eta$ (and $\epsilon$ for $\lambda=0$). The $r$-dependent B\nobreakdash-O transition amplitude, which we denote by $g_{\lambda,\eta}(j^\pi \to (j_1^{\pi_1}, j_2^{\pi_2}) j^\prime)$, causes the decays of double-heavy hadrons into pairs of heavy hadrons. The coupling potentials which mediate the decays can be expressed as sum of products of B\nobreakdash-O transition amplitudes $g_{\lambda,\eta}$, which can in principle be calculated using lattice QCD with two static color sources, and angular-momentum coefficients, which can be calculated analytically. This was shown in Ref.~\cite{Braa24} for the decays of hidden-heavy hadrons into pairs of heavy hadrons.

To calculate the coupling potential between a double-heavy hadron state and a pair of heavy hadrons, it is useful to express the latter in such a way that allows the B\nobreakdash-O symmetries to be exploited. The B\nobreakdash-O expansion in Eq.~\eqref{pairexpansion} uses the kets in Eq.~\eqref{stateHH} as a basis for the static-angular-momentum states of a heavy-hadron pair. An equivalent basis can be constructed by taking the direct product of the heavy-spin states in Eq.~\eqref{spinstate} and the light-QCD states in Eq.~\eqref{pairangularparity} and then expanding it into states with static-angular-momentum quantum numbers $S$ and $M_S$:
\begin{multline}
\vket{\Bigl[\Bigl(\tfrac{1}{2}^+(+), \tfrac{1}{2}^+(-)  \Bigr) S_Q, (j_1^{\pi_1}(+), j_2^{\pi_2}\Bigl(-)\Bigr) j^\prime \Bigr]}{S}{M_S}{} \\
= \sum_{m_{S_Q},\lambda} \cg{S_Q}{m_{S_Q}}{j^\prime}{\lambda}{S}{M_S} \vket{\Bigl(\tfrac{1}{2}^+(+), \tfrac{1}{2}^+(-)  \Bigr)}{S_Q}{m_{S_Q}}{} \vket{\Bigl(j_1^{\pi_1}(+), j_2^{\pi_2}(-)\Bigr)}{j^\prime}{\lambda}{}.
\label{stateHHalt}
\end{multline}
The advantage of this basis is that the spin of the $QQ$ pair, which is conserved in the heavy-quark limit, is factored from the light-QCD state of the static-hadron pair. The unitary change-of-basis transformation between the states in Eqs.~\eqref{stateHH} and \eqref{stateHHalt} is
\begin{multline}
\vket{\Bigl[(\tfrac{1}{2}^+, j_1^{\pi_1}  ) J_1(+), (\tfrac{1}{2}^+, j_2^{\pi_2}) J_2(-) \Bigr]}{S}{M_S}{}
= (-1)^{2 j_1} \sqrt{\tilde{J}_1 \tilde{J}_2} \\
\times \sum_{j^\prime, S_Q} \sqrt{\tilde{\jmath}^\prime \tilde{S}_Q}
\begin{Bmatrix}
\tfrac{1}{2}		& \tfrac{1}{2}				& S_Q	\\
j_1		& j_2	& j^\prime	\\
J_1				& J_2						& S	\\
\end{Bmatrix} 
\vket{\Bigl[\Bigl(\tfrac{1}{2}^+(+), \tfrac{1}{2}^+(-)  \Bigr) S_Q, (j_1^{\pi_1}(+), j_2^{\pi_2}(-)\Bigr) j^\prime \Bigr]}{S}{M_S}{},
\label{wignereq1}
\end{multline}
where $\tilde{J}=2J+1$ and $\left\{\begin{smallmatrix} j_1 & j_2 & j_3 \\ j_4 & j_5 & j_6 \\ j_7 & j_8 & j_9 \end{smallmatrix}\right\}$ is a Wigner $9j$ symbol. The sign $(-1)^{2j_1}$ comes from changing the order of the operators for the static hadron at $+\frac{1}{2}\bm{r}$ and that for the heavy quark at $-\frac{1}{2}\bm{r}$. Using Eq.~\eqref{wignereq1} and recoupling of angular momenta, we can rewrite the last line of the heavy-hadron-pair B\nobreakdash-O expansion in Eq.~\eqref{pairexpansion} as
\begin{align}
\sum_{m_{S},m_{L_Q^\prime}} &\cg{S}{M_S}{L_Q^\prime}{m_{L_Q^\prime}}{J}{M_J} \vket{\Bigl[(\tfrac{1}{2}^+, j_1^{\pi_1}  ) J_1(+), (\tfrac{1}{2}^+, j_2^{\pi_2}) J_2(-) \Bigr]}{S}{M_S}{} Y_{L_Q^\prime}^{m_{L_Q^\prime}}(\bm{\hat{r}}) \notag \\
=& (-1)^{2 j_1 + L_Q^\prime + J} \sqrt{\tilde{J}_1 \tilde{J}_2 \tilde{S}} \sum_{j^\prime, S_Q, L} (-1)^{j^\prime + S_Q} \sqrt{\tilde{\jmath}^\prime \tilde{S}_Q \tilde{L}}
\begin{Bmatrix}
S_Q	& j^\prime	& S 	\\
L_Q^\prime	& J 	& L 	\\
\end{Bmatrix}
\begin{Bmatrix}
\tfrac{1}{2}		& \tfrac{1}{2}				& S_Q	\\
j_1		& j_2	& j^\prime	\\
J_1				& J_2						& S	\\
\end{Bmatrix} \notag \\
&\times \sum_{m_{S_Q},M_L} \cg{S_Q}{m_{S_Q}}{L}{M_L}{J}{M_J} \vket{\Bigl(\tfrac{1}{2}^+(+), \tfrac{1}{2}^+(-)\Bigr)}{S_Q}{m_{S_Q}}{}\notag \\
&\times\sum_{\lambda,m_{L_Q^\prime}} \cg{j^\prime}{\lambda}{L_Q^\prime}{m_{L_Q^\prime}}{L}{M_L} \vket{\Bigl(j_1^{\pi_1}(+), j_2^{\pi_2}(-)\Bigr)}{j^\prime}{\lambda}{} Y_{L_Q^\prime}^{m_{L_Q^\prime}}(\bm{\hat{r}}).
\label{wignereq2}
\end{align}
Equation~\eqref{wignereq2} is particularly useful for calculating the coupling potentials between double-heavy hadrons and pairs of heavy hadrons, since it allows the conservation of the B\nobreakdash-O quantum numbers $\Lambda_\eta^\epsilon$, the B\nobreakdash-O angular momentum $L$, and the total heavy spin $S_Q$ to be exploited.

The radial coupling potentials can be calculated from Eq.~\eqref{wignereq2} using the techniques detailed in Ref.~\cite{Braa24} and references therein. They are expressed as
\begin{align}
V^{J^P}_{S_Q,L,\eta}\Bigl(j^\pi, &L_Q \to \bigl[(\tfrac{1}{2}^+, j_1^{\pi_1}) J_1, (\tfrac{1}{2}^+, j_2^{\pi_2}) J_2\bigr] S, L_Q^\prime\Bigr) \notag \\
=& N (-1)^{2 j_1 + S_Q + L_Q + J} \sqrt{\tilde{J}_1\tilde{J}_2 \tilde{S} \tilde{S}_Q \tilde{L}}
\times\sum_{j^\prime} 
(-1)^{j^\prime}
\sqrt{\tilde{\jmath}^\prime}
\begin{Bmatrix}
S_Q  & j^\prime & S \\
L_Q^\prime & J & L \\
\end{Bmatrix}
\begin{Bmatrix}
\frac{1}{2} & \frac{1}{2} & S_Q \\
j_1 & j_2 & j^\prime \\
J_1 & J_2 & S \\
\end{Bmatrix}\notag\\
&\times \sum_{\lambda}
\cg{j}{\lambda}{L}{-\lambda}{L_Q}{0}
\cg{j^\prime}{\lambda}{L}{-\lambda}{L_Q^\prime}{0}
g_{\lambda,\eta}(j^\pi \to (j_1^{\pi_1}, j_2^{\pi_2}) j^\prime).
\label{mixpot}
\end{align}
The superscript on $V$ is quantum numbers $J^P$ that are exactly conserved. The subscripts on $V$ are quantum numbers $S_Q$, $L$, and $\eta$ that are conserved in the heavy-quark limit. The normalization coefficient $N$ is 1 unless the two static hadrons $j_1^{\pi_1}$ and $j_2^{\pi_2}$ are identical.  Even in this case, $N$ differs from 1 only if $J_1\neq J_2$, in which case $N=\sqrt{2}$. This dependence of $N$ on the quantum numbers has been suppressed in Eq.~\eqref{mixpot}.

\subsection{Selection Rules}

The expression for the coupling potentials in Eq.~\eqref{mixpot} can be used to derive model-independent results for the decays of double-heavy hadrons into pairs of heavy hadrons, that is, B\nobreakdash-O selection rules and relative partial decay rates. These results were derived in Ref.~\cite{Braa24} for the decays of hidden-heavy hadrons into pairs of heavy hadrons. Since those results apply just as well to the decays of a double-heavy hadron into pairs of heavy hadrons, we review them here very briefly. 

The existence of nonzero B\nobreakdash-O transition amplitudes $g_{\lambda,\eta}$ in Eq.~\eqref{mixpot} requires the B\nobreakdash-O quantum numbers $\lambda$ and $\eta$ (and $\epsilon$) to be conserved. From conservation of $\Lambda=\lvert \lambda \rvert$, we have $\Lambda \leq j^\prime$. In the most general case, the value of $j^\prime$ ranges from $\lvert j_1 - j_2\rvert$ to $j_1 + j_2$ and we have the selection rule
\begin{equation}
\Lambda \leq j_1 + j_2.
\label{lambdaselrule}
\end{equation}
If the two static hadrons are identical, we have a more restrictive selection rule due to the simultaneous conservation of $\eta$ according to Eq.~\eqref{staticpairidenticaleta}. In this case, which requires $j_2^{\pi_2}=j_1^{\pi_1}$ and $i_2=i_1$, the value of $j^\prime$ ranges from $0$ to $2 j_1$. The maximum value of $j^\prime=2 j_1$ is allowed only if $\eta = (-1)^{I - 2 i_1}$. If $\eta = -(-1)^{I - 2 i_1}$, the maximum allowed value is $j^\prime=2 j_1 -1$ and Eq.~\eqref{lambdaselrule} gets replaced by
\begin{equation}
\Lambda \leq 2 j_1 -1.
\end{equation}
For the case $\lambda=0$, one has to take into account conservation of the B\nobreakdash-O quantum number $\epsilon$ according to Eq.~\eqref{pairreflection}. If the two static hadrons are identical, simultaneous conservation of $\eta$ and $\epsilon$ implies the selection rule
\begin{equation}
\eta\,\epsilon=(-1)^{I - 2i_1}.
\end{equation}

Because of the symmetry under reflections $\mathcal{R}$, which requires
\begin{equation}
g_{-\lambda,\eta}(j^\pi \to (j_1^{\pi_1}, j_2^{\pi_2}) j^\prime)
= 
\pi\,\pi_1\pi_2(-1)^{j - j^\prime} g_{\lambda,\eta}(j^\pi \to (j_1^{\pi_1}, j_2^{\pi_2}) j^\prime),
\end{equation}
and the symmetries of the Clebsch-Gordan coefficients, which require
\begin{equation}
\cg{j}{-\lambda}{L}{\lambda}{L_Q}{0}
\cg{j^\prime}{-\lambda}{L}{\lambda}{L_Q^\prime}{0}
 =
(-1)^{L_Q - L_Q^\prime + j^\prime - j}
\cg{j}{\lambda}{L}{-\lambda}{L_Q}{0}
\cg{j^\prime}{\lambda}{L}{-\lambda}{L_Q^\prime}{0},
\end{equation}
the sum over $\lambda$ in the last line of Eq.~\eqref{mixpot} is automatically zero unless
\begin{equation}
\pi \pi_1 \pi_2 (-1)^{L_Q - L_Q^\prime} = +1.
\label{reflectioncondition}
\end{equation}
Introducing the parities $P=\pi(-1)^{L_Q}$ of the double-heavy hadron and $P_1=\pi_1$ and $P_2=\pi_2$ of the two heavy hadrons, we see that Eq.~\eqref{reflectioncondition} implies the selection rule
\begin{equation}
P=P_1 P_2 (-1)^{L_Q^\prime},
\end{equation}
that is, conservation of parity.

The Wigner $6j$ and $9j$ symbols in the second line of Eq.~\eqref{mixpot} are automatically zero unless conservation of angular momentum is satisfied. This implies several triangle conditions for the angular momenta involved, many of them trivial. The nontrivial ones are a selection rule for the total heavy-hadron-pair spin $S$:
\begin{equation}
\lvert S_Q - j^\prime \rvert \leq S \leq S_Q + j^\prime
\end{equation}
and a selection rule for the heavy-hadron-pair orbital angular momentum $L_Q^\prime$:
\begin{equation}
\lvert L - j^\prime \rvert \leq L_Q^\prime \leq L + j^\prime.
\end{equation}

\subsection{Relative Partial Decay Rates}

In general, calculating the decay widths of a double-heavy hadron into pairs of heavy hadrons requires solving a coupled-channel Schr\"odinger equation. However, in simple cases it is possible to derive model-independent relative partial decay rates without solving any Schr\"odinger equation, thanks to the expansion of the coupling potentials in Eq.~\eqref{mixpot}. Note that the dependence of the coupling potentials on the distance $r$ enters only through the transition amplitudes $g_{\lambda,\eta}$. If the sums over $j^\prime$ and $\lambda$ reduce to a single term, then the transition amplitude $g_{\lambda,\eta}$ factors out of the coupling potentials between double-heavy hadrons in the same B\nobreakdash-O multiplet and pairs of heavy hadrons in the same HQSS multiplets. In this case, the ratio of the coupling potentials is a constant given by ratios of square-root factors and Wigner $3j$ and $6j$ symbols. If furthermore the masses of the decaying double-heavy hadrons are well above the corresponding hadron-pair thresholds, then the kinetic energies of the heavy hadrons are much larger than their spin splittings. In these cases, the square of the ratio of the coupling potentials, which is a rational number, equals the ratio of the partial decay rates per unit of phase space:
\begin{multline}
\mathrm{d}\Gamma\bigl(\Psi_{(L^P,S_{Q})J}\to (\Phi_{j_1^{\pi_1},J_{1}}, \Phi_{j_2^{\pi_2},J_{2}})_{L_Q^\prime}\bigr) / v^{2 L_Q^\prime + 1} \propto \\
N^2 \tilde{S}_{Q} \tilde{J}_{1}\tilde{J}_{2}
\sum_{S} \tilde{S}
\begin{Bmatrix}
S_{Q}  & j^\prime & S \\
L_Q^\prime & J & L \\
\end{Bmatrix}^2
\begin{Bmatrix}
\frac{1}{2} & \frac{1}{2} & S_Q \\
j_1 & j_2 & j^\prime \\
J_{1} & J_{2} & S \\
\end{Bmatrix}^2.
\label{ratioseq}
\end{multline}

In Sec.~\ref{exampledecays}, we give explicit results for the relative partial decay rates for conventional double-heavy baryons into pairs of heavy hadrons. The general expression for the relative partial decay rates, as well as a more thorough discussion of the conditions for its applicability, can be found in Sec.~III.C of Ref.~\cite{Braa24}.

\section{Examples}
\label{examplesection}

In this section, we discuss explicit applications of the general expressions discussed in this paper.  In Sec.~\ref{exampledoublebaryon}, we describe the lowest B\nobreakdash-O potentials and multiplets for conventional double-heavy baryons ($QQq$). In Sec.~\ref{exampletetraquark}, we describe the lowest B\nobreakdash-O potentials and multiplets for double-heavy tetraquark mesons ($QQ\bar{q}\bar{q}$). In Sec.~\ref{exampledecays}, we discuss the decays of a conventional double-heavy baryon into pairs of heavy hadrons ($QQq \to Q\bar{q}+Qqq$).

\subsection{Conventional Double-Heavy Baryons}
\label{exampledoublebaryon}

As the separation of the $QQ$ pair goes to zero, a light-QCD configuration of a low-lying $QQq$ baryon approaches that of a light quark $q$ in the presence of a static $\bm{3}^\ast$ color source at the origin, that is, a static meson. The energy differences between these $\bm{3}^\ast q$ configurations are approximated by the mass differences between the corresponding HQSS multiplets of $\bar{Q}q$ mesons.%
\footnote{\label{quote}Simply put, ``a heavy core is a heavy core'' \cite{Eic17}.}
Using this correspondence, one can infer the B\nobreakdash-O potentials for a conventional double-heavy baryon and their ordering as $r\to0$ from the spectrum of single-heavy mesons in QCD. Note that analogous arguments have been used for a long time in quark models for heavy baryons; see, for instance, Ref.~\cite{Kar09} and references therein.

\begin{table}
\caption{\label{heavymesons}Lowest three multiplets of nonstrange heavy mesons with the flavor of a light antiquark $(Q\bar{q})$ and their quantum numbers. For the corresponding heavy mesons with the flavor of a light quark $(\bar{Q}q)$, the parity superscript $\pi$ in $j^\pi$ is $+$ instead of $-$ for $S$-wave mesons and $-$ instead of $+$ for $P^-$- and $P^+$-wave mesons.}
\begin{ruledtabular}
\begin{tabular}{lllll}
Wave		& $j^\pi$		& I 			& $J^P$		& Meson		\\
\hline
$S$		& $\frac{1}{2}^-$	& $\frac{1}{2}$	& $0^-,1^-$		& $B,B^\ast$	\\
$P^-$		& $\frac{1}{2}^+$	& $\frac{1}{2}$	& $0^+,1^+$	& $B_0^\ast, B_1$	\\
$P^+$	& $\frac{3}{2}^+$	& $\frac{1}{2}$	& $1^+,2^+$	& $B_1^\prime, B_2^\ast$	\\
\end{tabular}
\end{ruledtabular}
\end{table}

The lowest three HQSS multiplets of nonstrange heavy mesons are listed in Table~\ref{heavymesons}. The lowest $\bar{Q}q$ mesons belong to the $S$-wave multiplet with $j^\pi=\frac{1}{2}^+$ and $I=\frac{1}{2}$. Since $j=\frac{1}{2}$ and $\Lambda\leq j$, the lowest B\nobreakdash-O potential has $\Lambda=\frac{1}{2}$. The $r\to0$ color state of the two sources being antisymmetric requires $\pi\eta=-1$; see Sec.~\ref{doubleheavysection}. Therefore, the static-meson parity $\pi=+1$ implies $\eta=-1$. The lowest B\nobreakdash-O potential then has quantum numbers $\Lambda_\eta=\frac{1}{2}_u$. The same reasoning can be applied to the other HQSS multiplets in Table~\ref{heavymesons}. For the $P^-$-wave multiplet with $j^\pi=\frac{1}{2}^-$, the B\nobreakdash-O potential has quantum numbers $\frac{1}{2}_g$. For the $P^+$-wave multiplet with $j^\pi=\frac{3}{2}^-$ the B\nobreakdash-O potentials have quantum numbers $\frac{1}{2}_g^\prime$ and $\frac{3}{2}_g$ and they become degenerate in the limit $r\to0$.%
\footnote{\label{etanote}These B\nobreakdash-O quantum numbers $\eta$ correspond to our convention of taking into account the fermionic nature of the heavy quarks in its spin state; see Sec.~\ref{staticlimitsubsection}. If adopting the alternative convention of taking the fermionic nature into account in the static color source, one should substitute $\eta\to-\eta$.}

\begin{figure}
\includegraphics{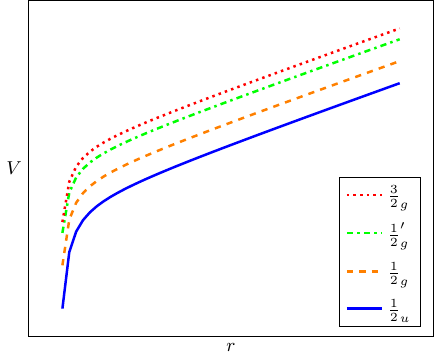}
\caption{\label{baryonpotentials}Pictorial representation of the lowest double-heavy-baryon potentials.}
\end{figure}

At short distances, the ${\frac{1}{2}}_u$, $\frac{1}{2}_g$, $\frac{1}{2}_g^\prime$, and $\frac{3}{2}_g$ potentials behave like attractive color-Coulomb potentials. At large distances, the ${\frac{1}{2}}_u$, $\frac{1}{2}_g$, $\frac{1}{2}_g^\prime$, and $\frac{3}{2}_g$ potentials increase linearly with $r$ \cite{Naj09}. These potentials are pictured in Fig.~\ref{baryonpotentials}. Note that the short-range ordering of the potentials mirrors the energy ordering of the corresponding static mesons.

\begin{table}
\caption{\label{doubleheavybaryons}Some of the lowest B-O multiplets of nonstrange double-heavy baryons with the flavor of a single light quark ($QQq$).}
\begin{ruledtabular}
\begin{tabular}{cllll}
Potential(s)		& I		& $L^P$ 		& $S_Q$ 		& $J^P$							\\
\hline
$\frac{1}{2}_u$	& $\frac{1}{2}$			& $\frac{1}{2}^+$	& 1		& $\frac{1}{2}^+, \frac{3}{2}^+$			\\
$\frac{1}{2}_u$	& $\frac{1}{2}$			& $\frac{1}{2}^-$	& 0		& $\frac{1}{2}^-$					\\
$\frac{1}{2}_u$	& $\frac{1}{2}$			& $\frac{3}{2}^-$	& 0		& $\frac{3}{2}^-$					\\
$\frac{1}{2}_u$	& $\frac{1}{2}$			& $\frac{3}{2}^+$	& 1		& $\frac{1}{2}^+,\frac{3}{2}^+,\frac{5}{2}^+$					\\
$\frac{1}{2}_u$	& $\frac{1}{2}$			& $\frac{5}{2}^+$	& 1		& $\frac{3}{2}^+,\frac{5}{2}^+,\frac{7}{2}^+$					\\[1em]
$\frac{1}{2}_g$	& $\frac{1}{2}$			& $\frac{1}{2}^-$	& 1		& $\frac{1}{2}^-, \frac{3}{2}^-$			\\
$\frac{1}{2}_g$	& $\frac{1}{2}$			& $\frac{1}{2}^+$	& 0		& $\frac{1}{2}^+$					\\
$\frac{1}{2}_g$	& $\frac{1}{2}$			& $\frac{3}{2}^+$	& 0		& $\frac{3}{2}^+$					\\[1em]
$\frac{1}{2}_g^\prime \text{ and } \frac{3}{2}_g$ & $\frac{1}{2}$ 	& $\frac{3}{2}^-$	& 1		& $\frac{1}{2}^-, \frac{3}{2}^-, \frac{5}{2}^-$	\\
$\frac{1}{2}_g^\prime$	& $\frac{1}{2}$		 	& $\frac{1}{2}^+$	& 0		& $\frac{1}{2}^+$					\\
$\frac{1}{2}_g^\prime\text{ and }\frac{3}{2}_g$ & $\frac{1}{2}$ 	& $\frac{3}{2}^+$	& 0		& $\frac{3}{2}^+$					\\
$\frac{1}{2}_g^\prime\text{ and }\frac{3}{2}_g$ & $\frac{1}{2}$	& $\frac{5}{2}^+$	& 0		& $\frac{5}{2}^+$					\\
\end{tabular}
\end{ruledtabular}
\end{table}

The lowest bound state in the ${\frac{1}{2}}_u$ potential has $L^P=\frac{1}{2}^+$; see Eq.~\eqref{lowestbomultiplet}. The B\nobreakdash-O exclusion principle in Eq.~\eqref{boexclusionprinciple} then requires $S_Q=1$. Therefore, the lowest conventional double-heavy baryons have $I=\frac{1}{2}$ and they form a B\nobreakdash-O multiplet with quantum numbers $J^P=\frac{1}{2}^+$ and $\frac{3}{2}^+$. The first two excited bound states in the ${\frac{1}{2}}_u$ potential have $L^P=\frac{1}{2}^-$ and $\frac{3}{2}^-$. The B\nobreakdash-O exclusion principle in Eq.~\eqref{boexclusionprinciple} then requires $S_Q=0$. Therefore, the first two excited conventional double-heavy baryons have $I=\frac{1}{2}$ and they form two separate B\nobreakdash-O singlets with $J^P=\frac{1}{2}^-$ and $J^P=\frac{3}{2}^-$. The same reasoning can be applied to determine the other B\nobreakdash-O multiplets of the ${\frac{1}{2}}_u$ potential as well as those in the higher double-heavy-baryon potentials. We list some of these double-heavy-baryon B\nobreakdash-O multiplets in Table~\ref{doubleheavybaryons}. 

\subsection{Double-Heavy Tetraquark Mesons}
\label{exampletetraquark}

As the separation of the heavy quarks pair goes to zero, a light-QCD configuration of a low-lying $QQ\bar{q}\bar{q}$ tetraquark meson approaches that of a $\bar{q}\bar{q}$ pair in the presence of a static $\bm{3}^\ast$ color source at the origin, that is, a static antibaryon. The energy differences between these $\bm{3}^\ast \bar{q}\bar{q}$ configurations are approximated by the mass differences between the corresponding HQSS multiplets of $\bar{Q}\bar{q}\bar{q}$ antibaryons.\footref{quote} Using this correspondence, one can infer the B\nobreakdash-O potentials for a double-heavy tetraquark meson and their ordering as $r\to0$ from the spectrum of single-heavy baryons in QCD. Note that analogous heavy-quark-symmetry arguments have been applied to double-heavy tetraquark mesons in Refs.~\cite{Eic17,Kar17}.

\begin{table}
\caption{\label{heavybaryons}Lowest two multiplets of nonstrange heavy baryons with the flavor of 2 light quarks $(Qqq)$ and their quantum numbers. For the corresponding heavy antibaryons with the flavor of 2 light antiquarks $(\bar{Q}\bar{q}\bar{q})$, the parity superscript $P$ in $J^P$ is $-$ instead of $+$.}
\begin{ruledtabular}
\begin{tabular}{lllll}
Wave		& $j^\pi$	& I 	& $J^P$					& Baryon		\\
\hline
$S^-$		& $0^+$	& $0$	& $\frac{1}{2}^+$				& $\Lambda_b$	\\
$S^+$		& $1^+$	& $1$	& $\frac{1}{2}^+,\frac{3}{2}^+$	& $\Sigma_b,\Sigma_b^\ast$	\\
\end{tabular}
\end{ruledtabular}
\end{table}

The lowest two HQSS multiplets of nonstrange heavy antibaryons are listed in Table~\ref{heavybaryons}. The lowest $\bar{Q}\bar{q}\bar{q}$ antibaryons belong to the $S$-wave multiplet with $j^\pi=0^+$ and $I=1$. Since $j=0$ and $\Lambda\leq j$, the lowest B\nobreakdash-O potential has $\Lambda=0$. The $r\to0$ color state of the two sources being antisymmetric requires $\pi\eta=-1$; see Sec.~\ref{doubleheavysection}. Therefore, $\pi=+1$ implies $\eta=-1$. From Eq.~\eqref{lightqcdreflection}, we have $\epsilon=+1$. The lowest B\nobreakdash-O potential then has quantum numbers $\Lambda_\eta=\Sigma_u^+$. The same reasoning can be applied to the other HQSS multiplet in Table~\ref{heavybaryons}. For the $S$-wave multiplet with $j^\pi=1^+$ and $I=1$, the B\nobreakdash-O potentials have quantum numbers $\Sigma_u^-$ and $\Pi_u$ and they become degenerate in the limit $r\to0$.\footref{etanote}

At small distances, the $\Sigma_u^+$, $\Sigma_u^-$, and $\Pi_u$ potentials behave like attractive color-Coulomb potentials. At large distances, the $\Sigma_u^+$, $\Sigma_u^-$, and $\Pi_u$ potentials approach the threshold for a pair of $S$-wave $Q\bar{q}$ mesons, as explained next.

As the separation of the heavy quarks grows larger, a light-QCD configuration of a $QQ\bar{q}\bar{q}$ tetraquark meson eventually approaches a pair of static $Q\bar{q}$ mesons. Thanks to this correspondence, one can use the spectrum of heavy mesons in QCD to infer the quantum numbers and the ordering of the B\nobreakdash-O potentials for heavy-meson pairs at large $r$. This correspondence was also illustrated in Ref.~\cite{Bic16} for pairs of $S$- or $P^-$-wave mesons.

The lowest nonstrange $Q\bar{q}$ mesons are those in the $S$-wave multiplet of Table~\ref{heavymesons}. So, the lowest double-heavy-tetraquark potentials at large $r$ approach twice the energy of an $S$-wave meson. A pair of $S$-wave mesons, with quantum numbers $j_1^{\pi_1}=j_2^{\pi_2}=\tfrac{1}{2}^-$ and $i_1=i_2=\tfrac{1}{2}$, can have total light-QCD angular-momentum $j^\prime=0$ and $1$ and total isospin $I=0$ or $1$. Since $\Lambda\leq j^\prime$, the heavy-meson-pair potentials have $\Lambda=0$ for $j^\prime=0$ and $\Lambda=0, 1$ for $j^\prime=1$.  As for the B\nobreakdash-O quantum number $\eta$, Eq.~\eqref{staticpairidenticaleta} for a pair of nonstrange identical static mesons requires
\begin{equation}
\eta =
\begin{cases}
(-1)^{j^\prime + 1}	& \text{if $I=0$,}	\\
(-1)^{j^\prime}		& \text{if $I=1$.}
\end{cases}
\end{equation}
Finally, the reflection quantum number for $\Lambda=0$ is $\epsilon=(-1)^{j^\prime}$. From these considerations, we see that lowest heavy-meson-pair potentials at large $r$ are:
\begin{enumerate}[(i)]
\item The $\Sigma_u^+$, $\Sigma_g^-$, and $\Pi_g$ potentials with $I=0$.
\item The $\Sigma_g^+$, $\Sigma_u^-$, and $\Pi_u$ potentials with $I=1$.
\end{enumerate}
These 6 potentials become degenerate at large $r$, where they approach twice the energy of an $S$-wave static meson.\footref{etanote}

As the $QQ$ separation goes to 0, a heavy-meson-pair potential connects to a double-heavy tetraquark potential with identical B\nobreakdash-O quantum numbers. As $r\to0$, the potential behaves like a repulsive color-Coulomb potential if $\eta=+1$ or like an attractive color-Coulomb potential if $\eta=-1$; see Eq.~\eqref{hadronsym}. So, the heavy-meson-pair potentials $\Sigma_g^-$ and $\Pi_g$ with $I=0$ and $\Sigma_g^+$ with $I=1$ connect to repulsive double-heavy tetraquark potentials. These correspond to the lowest states of a $\bar{q}\bar{q}$ pair bound to a static $\bm{6}$ color source at the origin, with quantum numbers $j^\pi=1^+$ for $I=0$ and $j^\pi=0^+$ for $I=1$. On the other hand, the heavy-meson-pair potentials $\Sigma_u^+$ with $I=0$ and $\Sigma_u^-$ and $\Pi_u$ with $I=1$ connect to the 3 attractive double-heavy tetraquark potentials discussed earlier. These correspond to the lowest states of a static antibaryon, with quantum numbers $j^\pi=0^+$ for $I=0$ and $j^\pi=1^+$ for $I=1$.

\begin{figure}
\includegraphics{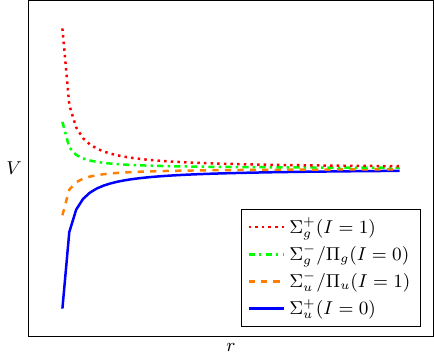}
\caption{\label{tetraquarkpotentials}Pictorial representation of the double-heavy tetraquark potentials approaching twice the energy of an $S$-wave meson at large $r$.}
\end{figure}

These potentials are pictured in Fig.~\ref{tetraquarkpotentials} (see also Ref.~\cite{Bic16}). Note that the pairs of potentials $(\Sigma_u^-,\Pi_u)$ and $(\Sigma_g^-,\Pi_g)$ are degenerate both as $r\to0$ and at large $r$. They have been represented using a single line for each pair to facilitate visualization. Also note that the short-range ordering of the attractive potentials $\Sigma_u^+$ and $\Sigma_u^-/\Pi_u$ mirrors the energy ordering of the corresponding static antibaryons. The short-range ordering of the repulsive potentials $\Sigma_g^+$ and $\Sigma_g^-/\Pi_g$ in Fig.~\ref{tetraquarkpotentials} is arbitrary, since the energy ordering of the corresponding $\bm{6}\bar{q}\bar{q}$ configurations is not known.

\begin{table}
\caption{\label{doubleheavytetraquarks}Some of the lowest B-O multiplets of nonstrange double-heavy tetraquark mesons with the flavor of two light antiquarks ($QQ\bar{q}\bar{q}$).}
\begin{ruledtabular}
\begin{tabular}{cllll}
Potential(s)		& I		& $L^P$ 		& $S_Q$ 		& $J^P$							\\
\hline
$\Sigma_u^+$	& $0$			& $0^+$	& 1		& $1^+$			\\
$\Sigma_u^+$	& $0$			& $1^-$	& 0		& $1^-$			\\
$\Sigma_u^+$	& $0$			& $2^+$	& 1		& $1^+,2^+,3^+$			\\[1em]
$\Sigma_u^- \text{ and } \Pi_u$	& $1$			& $1^+$	& 1		& $0^+,1^+,2^+$			\\
$\Sigma_u^-$	& $1$			& $0^-$	& 0		& $0^-$			\\
$\Pi_u$	& $1$			& $1^-$	& 0		& $1^-$			\\
$\Sigma_u^- \text{ and } \Pi_u$	& $1$			& $2^-$	& 0		& $2^-$			\\
\end{tabular}
\end{ruledtabular}
\end{table}

If the lowest double-heavy tetraquark potential $\Sigma_u^+$ admits bound states, then the lowest one has $L^P=0^+$; see Eq.~\eqref{lowestbomultiplet}. In this case, the B\nobreakdash-O exclusion principle in Eq.~\eqref{boexclusionprinciple} requires $S_Q=1$. Therefore, the lowest double-heavy tetraquark meson has quantum numbers $I=0$ and $J^P=1^+$, the same as the double-charm tetraquark meson $T_{cc}^+$ \cite{Aai22}. If the excited double-heavy tetraquark potentials $\Sigma_u^-$ and $\Pi_u$ admit bound states, then the lowest one has $L^P=1^+$. The B\nobreakdash-O exclusion principle in Eq.~\eqref{boexclusionprinciple} then requires $S_Q=1$. Therefore, the lowest double-heavy tetraquark mesons with $I=1$ form a B\nobreakdash-O multiplet with quantum numbers $J^P=0^+,1^+,2^+$. The same reasoning can be applied to determine other possible B\nobreakdash-O multiplets of double-heavy tetraquark mesons with higher energies. We list some of these double-heavy-tetraquark B\nobreakdash-O multiplets in Table~\ref{doubleheavytetraquarks}.

\subsection{Decays of a Double-Heavy Baryon into Heavy Hadrons}
\label{exampledecays}

Consider a static-hadron pair consisting of a static meson with the flavor of a light antiquark ($Q\bar{q}$), isospin $i_1$, and angular momentum and parity $j_1^{\pi_1}$, and a static baryon with the flavor of two light quarks ($Qqq$), isospin $i_2$, and angular momentum and parity $j_2^{\pi_2}$. The state of the pair can be expanded into states with total isospin and light-QCD angular-momentum quantum numbers $I=\lvert i_1 - i_2\rvert,\dotsc,i_1+i_2$ and $j^\prime=\lvert j_1 - j_2\rvert,\dotsc,j_1+j_2$. Each of these states corresponds to two distinct $r\to 0$ light-QCD multiplets with quantum number $\eta=-1$ or $+1$, respectively. Each multiplet contains $2 j^\prime + 1$ states with $\lambda$ ranging from $-j^\prime$ to $+j^\prime$, which become degenerate in the limit $r\to 0$. The corresponding B\nobreakdash-O potentials approach either an attractive or repulsive color-Coulomb potential in this limit, according to the rule in Eq.~\eqref{hadronsym}. We list in Table~\ref{mesbarnum} the isospin, the light-QCD angular-momentum, and the B\nobreakdash-O quantum numbers for pairs consisting of a static $Q\bar{q}$ meson from Table~\ref{heavymesons} and a static $Qqq$ baryon from Table~\ref{heavybaryons}.

\begin{table}
\caption{\label{mesbarnum}Isospin $I$, light-QCD angular-momentum $j^\prime$, and B-O quantum numbers $\Lambda_\eta$ for the pair consisting of a $S/P^-/P^+$-wave $Q\bar{q}$ meson and a $S^-/S^+$-wave $Qqq$ baryon.}
\begin{ruledtabular}
\begin{tabular}{rlll}
		& I 					& $j^\prime$	& $\Lambda_\eta$						\\
\hline
$S+S^-$	& $\frac{1}{2}$			& $\frac{1}{2}$				& $\frac{1}{2}_u, \frac{1}{2}_g$ 					\\
$P^-+S^-$	& $\frac{1}{2}$			& $\frac{1}{2}$				& $\frac{1}{2}_u, \frac{1}{2}_g$ 					\\
$P^++S^-$	& $\frac{1}{2}$			& $\frac{3}{2}$				& $\frac{1}{2}_u, \frac{1}{2}_g,\frac{3}{2}_u,\frac{3}{2}_g$ 	\\
$S+S^+$	& $\frac{1}{2},\frac{3}{2}$	& $\frac{1}{2},\frac{3}{2}$			& $\frac{1}{2}_u, \frac{1}{2}_g,\frac{3}{2}_u,\frac{3}{2}_g$ 	\\
$P^-+S^+$	& $\frac{1}{2},\frac{3}{2}$	& $\frac{1}{2},\frac{3}{2}$			& $\frac{1}{2}_u, \frac{1}{2}_g,\frac{3}{2}_u,\frac{3}{2}_g$ 	\\
$P^++S^+$	& $\frac{1}{2},\frac{3}{2}$	& $\frac{1}{2},\frac{3}{2},\frac{5}{2}$			& $\frac{1}{2}_u, \frac{1}{2}_g,\frac{3}{2}_u,\frac{3}{2}_g,\frac{5}{2}_u,\frac{5}{2}_g$ 	\\
\end{tabular}
\end{ruledtabular}
\end{table}

At large distances, the B\nobreakdash-O potentials of a static-meson-baryon pair approach a constant given by the sum of the energies of the static meson and the static baryon. On the other hand, the B\nobreakdash-O potentials of a conventional double-heavy baryon increase linearly at large $r$. Then a double-heavy-baryon potential that is attractive at small $r$ and a meson-baryon-pair potential with the same B\nobreakdash-O quantum numbers $\Lambda_\eta$ must necessarily have an avoided crossing at some intermediate $r$. The transition from a double-heavy baryon to a meson-baryon pair requires the creation of a light quark-antiquark pair, which is dynamically suppressed in low-energy QCD. This implies that the potentials have a \emph{narrow avoided crossing.}

As discussed in Sec.~\ref{exampledoublebaryon}, the lowest B\nobreakdash-O potentials of a conventional double-heavy baryon are $\frac{1}{2}_u$, $\frac{1}{2}_g$, $\frac{1}{2}_g^\prime$ and $\frac{3}{2}_g$. The B\nobreakdash-O quantum numbers $\frac{1}{2}_u$ and $\frac{1}{2}_g$ can be generated by all of the meson-baryon pairs listed in Table~\ref{mesbarnum}, so the $\frac{1}{2}_u$, $\frac{1}{2}_g$, $\frac{1}{2}_g^\prime$ double-heavy baryon potentials have narrow avoided crossings with the corresponding meson-baryon potentials. The B\nobreakdash-O quantum numbers $\frac{3}{2}_g$ can be generated by any of the meson-baryon pairs listed in Table~\ref{mesbarnum} except for $S+S^-$ and $P^-+S^-$, so the $\frac{3}{2}_g$ double-heavy baryon potential has narrow avoided crossings with the corresponding meson-baryon potentials.

Narrow avoided crossings require nonvanishing B\nobreakdash-O transition rates $g_{\lambda,\eta}$, which are responsible for the decays of conventional double-heavy baryons into a heavy meson and a heavy baryon. In simple cases where a single transition rate $g_{\lambda,\eta}$ dominates, it factors out of the ratio of the mixing potentials of double-heavy baryons from the same B\nobreakdash-O multiplet into a heavy meson and a heavy baryon from the same HQSS multiplets. One can then give model-independent predictions for relative partial decay rates using the B\nobreakdash-O symmetries only. These can be calculated using the same techniques explained in Ref.~\cite{Braa24} for the decays of hidden-heavy hadrons into pairs of heavy hadrons.

\begin{table}
\caption{\label{baryon1}Relative partial decay rates into a heavy meson and a heavy baryon for double-heavy baryons in the ground state $\frac{1}{2}_u$ potential with $I=\frac{1}{2}$, $L^P=\frac{1}{2}^+$, $S_Q=1$, and $J^{P}=\frac{1}{2}^+$ or $\frac{3}{2}^+$ . Decays into $S+S^+$ and $P^++S^+$ ($D$-wave) are not listed, because the relative partial decay rates are not completely determined by B-O symmetries. For relative partial decay rates into a charged $\Sigma_b^{(\ast)\pm}$ versus a neutral $\Sigma_b^{(\ast)0}$, multiply by an additional isospin factor of 2 in favor of the $\Sigma_b^{(\ast)\pm}$ final state.}
\begin{ruledtabular}
\begin{tabular}{lcrr}
& & $\frac{1}{2}^+$	& $\frac{3}{2}^+$	\\
\hline
\multirow{2}{*}{$S+S^-$ ($P$-wave)}	& $B \Lambda_b$ 		& 1 	& 4	\\
												& $B^\ast \Lambda_b$	& 11 	& 8	\\
\hline
\multirow{2}{*}{$P^-+S^-$ ($S$-wave)}	& $B_0^\ast \Lambda_b$ 	& 3 	& 0	\\
													& $B_1 \Lambda_b$			& 1 	& 4	\\
\hline
\multirow{2}{*}{$P^++S^-$ ($D$-wave)}	& $B_1^\prime \Lambda_b$ 			& 1 	& 4	\\
														& $B_2^\ast \Lambda_b$	& 7 	& 4	\\
														\hline
\multirow{4}{*}{$P^-+S^+$ ($S$-wave)}	& $B_0^\ast \Sigma_b$ 			& 3 	& 0	\\
														& $B_0^\ast \Sigma_b^\ast$	& 0 	& 12	\\
																												& $B_1 \Sigma_b$	& 25 	& 4	\\
																																										& $B_1 \Sigma_b^\ast$	& 8 	& 20	\\
															\hline
\multirow{4}{*}{$P^-+S^+$ ($D$-wave)}	& $B_0^\ast \Sigma_b$ 			& 0 	& 6	\\
														& $B_0^\ast \Sigma_b^\ast$	& 3 	& 6	\\
																												& $B_1 \Sigma_b$	& 4 	& 10	\\
																																										& $B_1 \Sigma_b^\ast$	& 29 	& 14	\\
\hline
\multirow{4}{*}{$P^++S^+$ ($S$-wave)}	& $B_1^\prime \Sigma_b$ 			& 8 	& 2	\\
														& $B_1^\prime \Sigma_b^\ast$	& 49 	& 10	\\
																												& $B_2^\ast \Sigma_b$	& 0 	& 30	\\
																																										& $B_2^\ast \Sigma_b^\ast$	& 15 	& 30	\\
\end{tabular}
\end{ruledtabular}
\end{table}

\begin{table}
\caption{\label{baryon2}Relative partial decay rates into a heavy meson and a heavy baryon for double-heavy baryons in the first excited $\frac{1}{2}_g$ potential with $I=\frac{1}{2}$, $L^P=\frac{1}{2}^-$, $S_Q=1$, and $J^{P}=\frac{1}{2}^-$ or $\frac{3}{2}^-$. Decays into $P^-+S^+$ and $P^++S^+$ ($P$-wave) are not listed, because the relative partial decay rates are not completely determined by B-O symmetries. For relative partial decay rates into a charged $\Sigma_b^{(\ast)\pm}$ versus a neutral $\Sigma_b^{(\ast)0}$, multiply by an additional isospin factor of 2 in favor of the $\Sigma_b^{(\ast)\pm}$ final state.}
\begin{ruledtabular}
\begin{tabular}{lcrr}
& & $\frac{1}{2}^-$	& $\frac{3}{2}^-$	\\
\hline
\multirow{2}{*}{$S+S^-$ ($S$-wave)}	& $B \Lambda_b$ 		& 3 	& 0	\\
												& $B^\ast \Lambda_b$	& 1 	& 4	\\
\hline
\multirow{2}{*}{$P^-+S^-$ ($P$-wave)}	& $B_0^\ast \Lambda_b$ 	& 1 	& 4	\\
													& $B_1 \Lambda_b$			& 11 	& 8	\\
\hline
\multirow{2}{*}{$P^++S^-$ ($P$-wave)}	& $B_1^\prime \Lambda_b$ 			& 19 	& 4	\\
														& $B_2^\ast \Lambda_b$	& 5 	& 20	\\
														
\hline
\multirow{4}{*}{$S+S^+$ ($S$-wave)}	& $B \Sigma_b$ 			& 3 	& 0	\\
														& $B \Sigma_b^\ast$	& 0 	& 12	\\
																												& $B^\ast \Sigma_b$	& 25 	& 4	\\
																																										& $B^\ast \Sigma_b^\ast$	& 8 	& 20	\\
\hline
\multirow{4}{*}{$S+S^+$ ($D$-wave)}	& $B \Sigma_b$ 			& 0 	& 6	\\
														& $B \Sigma_b^\ast$	& 3 	& 6	\\
																												& $B^\ast \Sigma_b$	& 4 	& 10	\\
																																										& $B^\ast \Sigma_b^\ast$	& 29 	& 14	\\
\hline
\multirow{4}{*}{$P^++S^+$ ($F$-wave)}	& $B_1^\prime \Sigma_b$ 			& 0 	& 18	\\
														& $B_1^\prime \Sigma_b^\ast$	& 9 	& 18	\\
																												& $B_2^\ast \Sigma_b$	& 8 	& 14	\\
																																										& $B_2^\ast \Sigma_b^\ast$	& 55 	& 22	\\
\end{tabular}
\end{ruledtabular}
\end{table}

\begin{table}
\caption{\label{baryon3}Relative partial decay rates into a heavy meson and a heavy baryon for double-heavy baryons in the coupled second and third excited $\frac{1}{2}_g^\prime$ and $\frac{3}{2}_g$ potentials with $I=\frac{1}{2}$, $L^P=\frac{3}{2}^-$, $S_Q=1$, and $J^{P}=\frac{1}{2}^-$, $\frac{3}{2}^-$, or $\frac{5}{2}^-$. Decays into $P^++S^-$, $S+S^+$, $P^-+S^+$, and $P^++S^+$ are not listed, because the relative partial decay rates are not completely determined by B-O symmetries.
}
\begin{ruledtabular}
\begin{tabular}{lcrrr}
& & $\frac{1}{2}^-$	& $\frac{3}{2}^-$	& $\frac{5}{2}^-$	\\
\hline
\multirow{2}{*}{$S+S^-$ ($D$-wave)}	& $B \Lambda_b$ 		& 0 	& 3	& 8	\\
							& $B^\ast \Lambda_b$	& 20 	& 17	& 12	\\
							\hline
\multirow{2}{*}{$P^-+S^-$ ($P$-wave)}	& $B_0^\ast \Lambda_b$ 	& 8 	& 5	& 0	\\
							& $B_1 \Lambda_b$		& 4 	& 7	& 12	\\
\end{tabular}
\end{ruledtabular}
\end{table}

The relative partial decay rates of double-heavy baryons in B\nobreakdash-O multiplets with the same quantum numbers as the ground states of the $\frac{1}{2}_u$ potential, the $\frac{1}{2}_g$ potential, and the $\frac{1}{2}_g^\prime$ and $\frac{3}{2}_g$ potentials are listed in Tables~\ref{baryon1}, \ref{baryon2}, and \ref{baryon3}, respectively. Note that these relative partial decay rates take into account only the angular-momentum factors from Eq.~\eqref{ratioseq} and not the relevant isospin Clebsch-Gordan coefficients. The square of the isospin coefficients factors out in all cases listed in the tables except for relative partial decay rates into a charged heavy baryon, $\Sigma_b^{(\ast)\pm}$, versus a neutral heavy baryon, $\Sigma_b^{(\ast)0}$. Only in these cases, isospin symmetry requires a multiplicative factor of 2 in favor of the final state with a $\Sigma_b^{(\ast)\pm}$.

\section{Conclusions}
\label{conclusionsection}

In this work, we have used the diabatic Born-Oppenheimer (B\nobreakdash-O) approximation for QCD to derive model-independent predictions for the decays of double-heavy hadrons into pairs of heavy hadrons. The coupling potential between a double-heavy hadron and a heavy-hadron pair is expressed in Eq.~\eqref{mixpot} as the sum of products of angular-momentum coefficients and B\nobreakdash-O transition amplitudes. This expression is analogous to that for the coupling potential between a hidden-heavy hadron and a pair of heavy hadrons derived in Ref.~\cite{Braa24}. Unlike the hidden-heavy hadron case of Ref.~\cite{Braa24}, however, the double-heavy case presents additional complications due to the prominent role played by identical particles and isospin. We have derived the B\nobreakdash-O exclusion principle for double-heavy hadrons, given by Eq.~\eqref{boexclusionprinciple}, and shown that it agrees with quark models. We have shown how the B\nobreakdash-O symmetries of a pair of heavy hadrons are related to the parities, spins, and flavors of the two individual hadrons. We have pointed out that a hadron-pair potential necessarily connects to a color-Coulomb potential at small $r$. We have derived a general rule in Eq.~\eqref{hadronsym} that determines whether this color-Coulomb potential is attractive or repulsive. This rule had been deduced previously based on lattice-QCD calculations of specific cases \cite{Bic16}. We have derived the exclusion principles from identical heavy hadrons, given by Eq.~\eqref{pairexclusionprinciple}, and shown that it agrees with quark models.

In addition to the general expressions, we have examined several concrete applications. We determined the lowest B\nobreakdash-O multiplets for conventional double-heavy baryons and for double-heavy tetraquark mesons. We also derived relative partial decay rates for the decays of a double-heavy baryon into a heavy meson and a heavy baryon. The method used to derive these relative partial decay rates can be easily generalized to any double-heavy hadron and to any pair of heavy hadrons, provided knowledge of the B\nobreakdash-O quantum numbers of the double-heavy hadron and the light-QCD quantum numbers of the two heavy hadrons.

Finally, let us note that the B\nobreakdash-O transition amplitudes $g_{\lambda,\eta}$ can in principle be calculated \textit{ab initio} using lattice QCD with two static color sources. Plugging these transition amplitudes into Eq.~\eqref{mixpot} yields mixing potentials between double-heavy hadrons and heavy-hadron pairs that are completely determined by lattice QCD and B\nobreakdash-O symmetries. These mixing potentials can then be used to calculate double-heavy-hadron masses and decay widths by solving a multichannel Schr\"odinger equation with coupled double-heavy-hadron and heavy-hadron-pair channels.

\acknowledgments{%
I acknowledge valuable discussions with E.~Braaten, who first pointed out to me that heavy-hadron-pair potentials at long range must be connected to double-heavy-hadron potentials at short range. This research was supported by the U.S. Department of Energy under Grant No.\ DE-SC0011726. This work contributes to the goals of the US DOE ExoHad Topical Collaboration, Contract No.\ DE-SC0023598.%
}

\bibliography{doubleheavybib}

\end{document}